\documentclass[aps, prb, twocolumn, showpacs, superscriptaddress]{revtex4-1}

\usepackage{amsmath,amssymb}
\usepackage{graphicx}
\usepackage{dcolumn}
\usepackage{bm}
\begin{document}

\title {Topological quantum phase transition in Kane-Mele-Kondo lattice model}
\author{Yin Zhong}
\email{zhongy05@hotmail.com}
\affiliation{Center for Interdisciplinary Studies $\&$ Key Laboratory for
Magnetism and Magnetic Materials of the MoE, Lanzhou University, Lanzhou 730000, China}
\author{Yu-Feng Wang}
\affiliation{Center for Interdisciplinary Studies $\&$ Key Laboratory for
Magnetism and Magnetic Materials of the MoE, Lanzhou University, Lanzhou 730000, China}
\author{Han-Tao Lu}
\affiliation{Center for Interdisciplinary Studies $\&$ Key Laboratory for
Magnetism and Magnetic Materials of the MoE, Lanzhou University, Lanzhou 730000, China}
\author{Hong-Gang Luo}
\email{luohg@lzu.edu.cn}
\affiliation{Center for Interdisciplinary Studies $\&$ Key Laboratory for
Magnetism and Magnetic Materials of the MoE, Lanzhou University, Lanzhou 730000, China}
\affiliation{Beijing Computational Science Research Center, Beijing 100084, China}

\date{\today}

\begin{abstract}
We systematically explore the ground-state phase diagram of the Kane-Mele-Kondo lattice model on the honeycomb lattice, particularly, we focus on its magnetic properties, which are ubiquitous and important in many realistic heavy fermion materials but have not been clearly studied in the previous publication[Feng, Dai, Chung, and Si, Phys. Rev. Lett. \textbf{111}, 016402 (2013)]. Beside the Kondo insulator found in that paper, two kinds of antiferromagnetic spin-density-wave phases are discovered. One is the normal antiferromagnetic spin-density-wave state and the other is a nontrivial topological antiferromagnetic spin-density-wave state, which shows the unexpected $Z_{2}$ topological order with a quantized spin Hall conductance and a helical edge-state similar to the hotly studied topological insulators. The quantum spin Hall insulator in Feng \text{et al.}'s work is found to be absent since it is always unstable to antiferromagnetic spin-density-wave states at least at the mean-field level in our model. Interestingly, the transition between the two spin-density-wave phases is an exotic topological quantum phase transition, whose critical behavior is described by an emergent three-dimensional quantum electrodynamics, in which conduction electrons contribute to the low-energy Dirac fermions while the spin-wave fluctuation of local spins gives rise to an effective dynamic U(1) gauge-field. Such nontrivial transition shows radical critical thermodynamic, transport and single-particle behaviors, which provide a fingerprint for this transition. Additionally, the transition of antiferromagnetic spin-density-wave states to the Kondo insulator is found to be first-order. The introduction of two novel magnetic phases and their topological quantum phase transition show rich and intrinsic physics involving in the Kane-Mele-Kondo lattice model.
\end{abstract}

\maketitle

\section{Introduction} \label{intr}
The interplay of the strong electron correlation and the band topology attracts much attention in recent years after the discovery of the two dimensional quantum spin Hall insulator (QSH) and three-dimensional topological band insulator.\cite{Bernevig2006,Haldane,Kane2006,Wiedmann,Shen2009,Hasan2010,Qi2011} Both of these two novel insulators are the bulk band insulator but have nontrivial gapless edge or surface states protected by the time-reversal symmetry.\cite{Qi2011} When strongly electron correlated effect is considered, understanding the fate of the bulk topology and the corresponding boundary states is still a challenge in the condensed matter community.\cite{Raghu2008,Ran2008,Qi2008,Levin2009,Regnault,Neupert,Mei,Sun,Sheng2011,Ruegg2012,Zhong2013b,Ruegg2013,Dzero,Dzero2012,Kim2012,Mong,Essin,He2011,He2012,Zhong2013,Yoshida2013b,Rachel,Hohenadler2011,Zheng2011,Li2011,Imada2011,Wu2012,Feng,Yoshida,Assaad2013} Despite of the hardness of this issue, many new and exotic states of matter have been uncovered, e.g., the topological
Mott insulator,\cite{Raghu2008} fractional Chern/topological insulator,\cite{Ran2008,Qi2008,Levin2009,Regnault,Neupert,Mei,Sun,Sheng2011} Z$_{2}$-fractionalized Chern/topological insulator,\cite{Ruegg2012,Zhong2013b,Ruegg2013} topological Kondo insulator\cite{Dzero,Dzero2012,Kim2012} and topological antiferromagnetic spin-density-wave phase.\cite{Mong,Essin,He2011,He2012,Zhong2013,Yoshida2013b}

Among them, the concept of topological Kondo insulator (TKI) has been proposed as an extension of the original
topological (band) insulators into the heavy fermion systems.\cite{Dzero} In usual topological
insulators the spin-orbit coupling is encoded in a spin-dependent hopping amplitudes between different unit cells. The TKI is produced by the spin-orbit coupling associated with the hybridization between conduction and local electrons and its resulting energy band is also heavily renormalized by the local strong correlation of local electrons.\cite{Dzero,Dzero2012,Kim2012} In comparison
to the conventional Kondo insulator (KI),\cite{Aeppli,Tsunetsugu,Riseborough} the TKI remains a renormalized band insulator but it possesses certain conducting surface-states, which could be identified experimentally from the usual KI compounds. Theoretically, the three-dimensional TKI may be relevant to certain heavy fermion compounds with strong spin-orbit coupling. Very recent experimental studies on the classic Kondo insulator SmB$_{6}$ seem to support that
SmB$_{6}$ is a putative three-dimensional topological Kondo insulator.\cite{NXu2013,Neupane2013,Li2013} This motivates us further study of the TKI in these strongly correlated systems.

Very recently, a microscopic Kondo lattice model on the two-dimensional honeycomb lattice with spin-orbit coupling between conduction electrons has been proposed to understand the interplay between spin-orbit coupling (bulk topology) and Kondo screening effect in two-dimensional system. \cite{Feng} In that paper the authors found the KI and the QSH insulators paramagnetic phases, and identified continuous second-order quantum phase transition between these two phases.

However, it should be noted that the magnetic long-range ordered states, which commonly appear and rather relevant for both the theoretical Kondo lattice-like models and realistic heavy fermion compounds, have not been clearly explored in the previous work of Feng \text{et al.}\cite{Feng}. As a matter fact, in some heavy fermion materials, the antiferromagnetism is even near by or coexistent with unconventional superconductivity. In this case the magnetic fluctuation is believed to be driving force of such superconducting instability.\cite{Pfleiderer} Furthermore, many experimentally observed exotic non-Fermi liquid behaviors are also close to the magnetic instability and its related quantum critical points.\cite{Rosch} On the other hand, intuitively, based on our previous studies on the kondo lattice models on the honeycomb lattice,\cite{Zhong2012b,Zhong2013} we suspect that the paramagnetic states are not stable to magnetic phases, particularly, the QSH insulator found by Feng \text{et al.} is always unstable to the antiferromagnetic spin-wave-density (SDW) states if the Kondo coupling between local spins and conduction electrons is not vanishing small. Another interesting feature of including the magnetic order is that it could give rise to an example of the mentioned topological antiferromagnetic SDW (T-SDW) phase.\cite{He2011,He2012,Zhong2013,Yoshida2013b} The T-SDW phase is identified as an antiferromagnetic SDW state with some nontrivial bulk topological quantum numbers or bulk charge/spin response. This state is indeed beyond the classic Landau's symmetry-breaking classification of matter since the T-SDW and the normal SDW (N-SDW) states have identical bulk symmetry (breaking the spin-rotation invariance due to the antiferromagnetic order).\cite{Wen2004}
In some sense, the T-SDW states can be considered as an interesting extension of the usual topological insulator with the broken time-reversal symmetry by the magnetic long-range order.

In the present work, we discuss these interesting issues in the Kane-Mele-Kondo lattice model proposed in Ref.\onlinecite{Feng} at half-filling. To be specific, we will use the mean-field decoupling to obtain magnetic T and N-SDW phases as well as the paramagnetic KI and QSH insulator phases. Importantly, the magnetic ordered T and N-SDW states are novel features of the current work beyond that in the previous paper of Feng \text{et al.}\cite{Feng}. We compare ground-state energies of those phases to establish the global phase diagram for the Kane-Mele-Kondo lattice model. After identifying those phases, the corresponding topological Chern number is explicitly calculated and the Chern-Simons and quantum electrodynamics-like effective field theories are utilized to capture the possible topological properties by taking important quantum fluctuations around the mean-field solutions into account.

It is found that the low energy effective theory of the $Z_{2}$ topological ordered T-SDW phase can be constructed by the hydrodynamics approach, which is successfully applied to the well-known integer and fractional quantum Hall effect.\cite{Wen2004} The hydrodynamics approach gives rise to a mutual U(1)$\times$ U(1) Chern-Simons action coupled to the external electromagnetic and spin gauge-field. Furthermore, the bulk-boundary correspondence of the Chern-Simons theory provides a candidate of the helical edge-state for the T-SDW phase. More interestingly, the topological quantum phase transition between the T and N-SDW phases is described by the unexpected three-dimensional quantum electrodynamics, which appears in many important condensed theory issues, e.g. the gauge theory description of high-temperature supconductivity,\cite{Wen} the quantum spin liquids in quantum magnets\cite{Sachdev2011} and the deconfined quantum criticality\cite{Senthil-dec}. Such emergent quantum electrodynamics is a novel feature of the present work and is also not covered in Ref.\onlinecite{Feng}. In the present case, we find that conduction electrons contribute to the low-energy Dirac fermions while the spin-wave fluctuation gives rise to an effective dynamic U(1) gauge-field. Critical quasiparticle, thermodynamic and transport behaviors of the emergent quantum electrodynamics are analyzed at the lowest-order and the one-loop levels.

Furthermore, we find that if the Kondo coupling is turned on, the QSH insulator becomes unstable in comparison to the antiferromagnetic SDW states, thus at least within framework of our mean-field theory, the ground-state phase diagram will not cover the QSH state found in Ref.\onlinecite{Feng} and only T/N-SDW and KI phases remain. The transition of these two magnetic states to the KI phase is first-order. Additionally, we have also discussed the doped Kane-Mele-Kondo lattice model and ferromagnetism may be relevant when the system is deviated from the half-filling case. The relation to the TKI and many implications from both numerical and experimental studies are also argued with an aim toward the realization of the Kane-Mele-Kondo model and confirmation of the various phases of the ground-state phase diagram.

To briefly summarize, we focus on the magnetic properties of the Kane-Mele-Kondo lattice model, and find two kinds of novel ordered phases, namely, the N- and T-SDW phases. Furthermore, the topological quantum phase transition between two magnetic phases found is a novel featu\emph{}re for this model, and its effective field theory is the celebrated 2+1D QED, which appears in many important and interesting issues, for example, deconfined  quantum criticality, gauge-theory description of the high temperature superconductivity and quantum spin liquids in quantum magnets. Finally, we also systematically discussed the topological properties of the model including topological Chern number, effective Chern-Simons field-theory for the Z$_2$ topological ordered phase, the edge-state Hamiltonian, and so on. These results have not clearly identified in the previous work published by Feng \textit{et al.} \cite{Feng}

The remainder of this paper is organized as follows. In Sec. \ref{sec2}, Kane-Mele-Kondo lattice model is introduced and its mean-field Hamiltonian is derived. In Sec. \ref{sec3}, we present detailed study of antiferromagnetic ordered phases. The T and N-SDW states are uncovered and their physical properties are inspected from the single particle band to topological properties. In Sec. \ref{sec4}, fluctuation correction is considered for the antiferromagnetic phases and effective field theories are derived and analyzed. We construct the ground-state phase diagram of Kane-Mele-Kondo lattice model in Sec.\ref{sec5}. Sec. \ref{sec6} and \ref{sec7} give some important discussions and extensions. Finally, Sec. \ref{sec8} is devoted to a brief conclusion.

\section{The Kane-Mele-Kondo lattice model}\label{sec2}
The following model we considered is the anisotropic Kondo lattice model on the honeycomb lattice at half-filling,
\begin{eqnarray}
&&H=H_{\text{K}}+H_{\parallel}+H_{\perp},\nonumber\\
&&H_{\text{K}}=-t\sum_{\langle ij\rangle \sigma}c_{i\sigma}^{\dag}c_{j\sigma}-t'\sum_{\langle\langle ij\rangle\rangle \sigma}\sigma e^{i\varphi_{ij}}c_{i\sigma}^{\dag}c_{j\sigma},\nonumber\\
&&H_{\parallel}=\frac{J_{\parallel}}{4}\sum_{i}(c_{i\uparrow}^{\dag}c_{i\uparrow}-c_{i\downarrow}^{\dag}c_{i\downarrow})(d_{i\uparrow}^{\dag}d_{i\uparrow}-d_{i\downarrow}^{\dag}d_{i\downarrow}),\nonumber\\
&&H_{\perp}=\frac{J_{\perp}}{2}\sum_{i}(c_{i\uparrow}^{\dag}c_{i\downarrow}d_{i\downarrow}^{\dag}d_{i\uparrow}+c_{i\downarrow}^{\dag}c_{i\uparrow}d_{i\uparrow}^{\dag}d_{i\downarrow}), \label{eq1}
\end{eqnarray}
where $H_{\text{K}}$ is the well-known Kane-Mele Hamiltonian,\cite{Kane2006} which supports the quantum spin Hall (QSH) effect with helical edge states (Right and left movers are locked with the up and down spin, respectively), $t$ and $t'$ are the nearest-neighbor and the next-nearest-neighbor hopping, respectively.\cite{Qi2011} The phase $\varphi_{ij}=\pm\frac{1}{2}\pi$ is introduced to give rise to a quantum spin Hall effect without external magnetic fields (the so-called QSH) and the positive phase is gained with anticlockwise hopping. Besides, the pseudofermion representation for local spins has been utilized as $S_{i}^{\alpha}=\frac{1}{2}\sum_{\sigma\sigma'}d_{i\sigma}^{\dag}\tau_{\sigma\sigma'}^{\alpha}d_{i\sigma'}$ with $\tau^{\alpha}$ being usual Pauli matrix and a local constraint $d_{i\uparrow}^{\dag}d_{i\uparrow}+d_{i\downarrow}^{\dag}d_{i\downarrow}=1$ enforced in each site. $H_{\parallel}$ denotes the magnetic instability due to the polarization of conduction electrons by local spins while $H_{\perp}$ describes the local Kondo screening effect resulting from spin-flip scattering process of conduction electrons by local moments.

\subsection{Some background}
The interplay between the Kondo screening effect and the magnetic instability on the honeycomb lattice without the nontrivial next-nearest-neighbor hopping term ($t'$) has been studied by some of the present authors in the previous work. \cite{Zhong2012b} In that work, we find either a direct first-order transition or a possible coexistence of the Kondo insulator and the N-SDW state by using the extended mean-field decoupling.
Furthermore, we have also discussed the Haldane-Kondo lattice model where the conduction electrons are described by the spinful Haldane model.\cite{Zhong2013} We found an extra antiferromagnetic phase, namely the T-SDW state, which has a quantized charge Hall conductance coexistent the magnetic order. The transition of T-SDW state into the conventional N-SDW state is found to fall into a XY-like topological quantum phase transition.

The model we study here is different to the Haldane-Kondo lattice model and more richer physics has been found in comparison to the previous one. In particular, the T-SDW in the present paper is different to the one in Ref.\onlinecite{Zhong2013} since the current T-SDW phase has a quantized spin Hall conductance and the edge-state is the helical edge-state. We now use the hydrodynamics approach to establish an effective theory for T-SDW phase by extracting the bulk topological properties and to identify existence of the non-chiral edge-state. More interestingly, the phase transition between T and N-SDW is now described by the three-dimensional quantum electrodynamics (QED-3) in contrast to the XY transition in the previous work. The QED-3 theory may give rise to a nontrivial universal conductance, which is also different to the previous work. In addition, we also provide many discussion on the doped system, which is important but not considered before.

Recently, Feng \textit{et al.} have studied the above model Eq. \ref{eq1} by using the slave boson mean-field theory. They mainly focus on the possible paramagnetic states, i.e., the Kondo insulator and the quantum spin hall insulator (two dimensional topological insulator). They found a second-order transition between those two paramagnetic states. However, the magnetic ordered states, which commonly appear in the Kondo lattice-like models and realistic heavy fermion compounds, are not studied. Intuitively, based on our previous studies on the kondo lattice models on the honeycomb lattice, we suspect that the paramagnetic states are not stable to magnetic phases, particularly, the quantum spin Hall insulator is always unstable to the antiferromagnetic SDW states if the Kondo coupling is not vanishing small. In the present paper we will discuss this issue in the following sections
and we will explore the various transitions among those magnetic and paramagnetic phases and establish the global phase diagram for the ground-state of the Kane-Mele-Kondo lattice model.

\subsection{Remark on the T-SDW state}
In the present paper, the T-SDW state is an antiferromagnetic long-range ordered state with the quantized spin Hall effect and a helical edge-state. Similar T-SDW phase seems to be found in a numerical way and the time-reversal symmetry may be recovered in the edge, thus the helical edge-state is protected by the recovered symmetry.\cite{Yoshida2013b} In contrast, in Ref.\onlinecite{He2011,Zhong2013}, the T-SDW phase represents a time-reversal symmetry-breaking antiferromagnetic state with the quantized charge Hall effect. The broken time-reversal symmetry leads to two chiral edge-states in the latter T-SDW phase, which is similar to the case of usual electric or bosonic integer quantum Hall effect where the external/effecitve large magnetic field explicitly breaks the time-reversal invariance and leads to energy bands with nontrivial Chern number.\cite{Wen2004,Senthil2012}

\subsection{Mean-field Hamiltonian of Kane-Mele-Kondo lattice model}
In this subsection, we present the mean-field decoupling and the consultant mean-field Hamiltonian, which is our starting point to study the ground-state phase diagram of the Kane-Mele-Kondo lattice model [Eq. (\ref{eq1})].
Although the mean-field treatment of generic two-dimensional lattice models needs caution in strong correlation scenario, it may provide qualitatively correct physics if quantum fluctuations are weak. In the following discussion,
we assume that the mean-field treatment is able to capture the basic features of the Kane-Mele-Kondo lattice model. Further study on this model in terms of advanced numerical techniques is desirable.

By using the mean-field decoupling as presented in Refs.\onlinecite{Zhong2013} and \onlinecite{Zhang2000} for the longitudinal and transverse interaction term $H_{\parallel}$, $H_{\perp}$, respectively, the resultant mean-field Hamiltonian reads
\begin{eqnarray}
&&H_{\text{MF}}=H_{\text{K}}+H_{\parallel}^{\text{MF}}+H_{\perp}^{\text{MF}}+E_{0},\nonumber\\
&&H_{\parallel}^{\text{MF}}=\frac{J_{\parallel}}{2}\sum_{k\sigma}[\sigma(-m_{c}d_{kA\sigma}^{\dag}d_{kA\sigma}+m_{d}c_{kA\sigma}^{\dag}c_{kA\sigma})\nonumber\\
&& \hspace{2cm} - (A\rightarrow B)], \nonumber\\
&&H_{\perp}^{\text{MF}}=\frac{J_{\perp}V}{2}\sum_{k\sigma}(c_{kA\sigma}^{\dag}d_{kA\sigma}+c_{kB\sigma}^{\dag}d_{kB\sigma}+h.c.),\nonumber\\
&&E_{0}=N_{s}(2J_{\parallel}m_{d}m_{c}+J_{\perp}V^{2}), \label{eq2}
\end{eqnarray}
where $N_s$ is the number of lattice sites. Several mean-field parameters have been defined as $\langle d_{iA\uparrow}^{\dag}d_{iA\uparrow}-d_{iA\downarrow}^{\dag}d_{iA\downarrow}\rangle = 2m_{d}$,
$\langle d_{iB\uparrow}^{\dag}d_{iB\uparrow}-d_{iB\downarrow}^{\dag}d_{iB\downarrow}\rangle = -2m_{d}$, $\langle c_{iA\uparrow}^{\dag}c_{iA\uparrow}-c_{iA\downarrow}^{\dag}c_{iA\downarrow}\rangle = -2m_{c}$,
$\langle c_{iB\uparrow}^{\dag}c_{iB\uparrow}-c_{iB\downarrow}^{\dag}c_{iB\downarrow}\rangle = 2m_{c}$ and
$-V=\langle c_{i\uparrow}^{\dag}d_{i\uparrow}+d_{i\downarrow}^{\dag}c_{i\downarrow}\rangle =
\langle c_{i\downarrow}^{\dag}d_{i\downarrow}+d_{i\uparrow}^{\dag}c_{i\uparrow}\rangle$. The mean-field parameters $m_{d}, m_{c}$ correspond to magnetization of local spins and conduction electrons, respectively, and
serve as usual magnetic order parameters. The non-vanishing Kondo hybridizing parameter $V$ denotes the onset of Kondo screening effect and it can also be considered as an order parameter although the formation of
Kondo screening does not imply any global symmetry-breaking. Besides, since we are considering a half-filled lattice, the local constraint of the pseudofermion has been safely neglected at the present mean-field level with chemical potential set to zero.\cite{Zhang2000}

With the mean-field Hamiltonian Eq. (\ref{eq2}) in hand, we can extract some but not all possible physical phases. Firstly, when all mean-field parameters are zero, the Hamiltonian will only have the conduction electron part and this just corresponds to the quantum spin Hall state.\cite{Kane2006} Secondly, when Kondo hybridizing parameter $V$ is non-vanishing but $m_{c}=m_{d}=0$, this is the Kondo state. Since the system is half-filling, one may expect that it is the Kondo insulating state.\cite{Tsunetsugu} Thirdly, if we have $m_{c},m_{d}\neq0$ with $V=0$, the corresponding state is the antiferromagnetic ordered state. Finally, if all of the mean-field parameters are nonzero, one may find a state where the antiferromagnetic long-range order coexists with the Kondo screening.\cite{Zhong2012b,Zhang2000} Interestingly, we will find an antiferromagnetic ordered state with a quantized spin Hall conductance, which is beyond our naive inspect due to blindness for the subtle topological properties. Detailed study on these possible states is the purpose of the present paper but the coexistent state will be left for future study due to its complexity and subtlety.

Before moving to the discussion on the possible states of Kane-Mele-Kondo lattice model in the next section, it is helpful to give a brief argument on the basic properties of Kane-Mele model and its corresponding Chern-Simons treatment since these issues may not appear in the literature of heavy fermions and the same technique will be used in the next section. Another motivation to do this is that when the mean-field order parameters are all zero, the Hamiltonian Eq. (\ref{eq2}) is reduced to the pure Kane-Mele model, which is just the quantum spin Hall insulator found in Ref. \onlinecite{Feng}.

\subsection{The Kane-Mele model}
In this subsection, we give a brief discussion about the basic property of the Kane-Mele model \cite{Kane2006}
\begin{eqnarray}
H_{\text{K}}=-t\sum_{\langle ij\rangle \sigma}c_{i\sigma}^{\dag}c_{j\sigma}-t'\sum_{\langle\langle ij\rangle\rangle \sigma}\sigma e^{i\varphi_{ij}}c_{i\sigma}^{\dag}c_{j\sigma}. \label{eq3}
\end{eqnarray}
It is useful to rewrite this single-particle Hamiltonian in the momentum space as
\begin{eqnarray}
H_{\text{K}}&&=\sum_{k\sigma}-t[f(k)c_{kA\sigma}^{\dag}c_{kB\sigma}+f^{\star}(k)c_{kB\sigma}^{\dag}c_{kA\sigma}] \nonumber\\
&& + 2t'\sigma\gamma(k)[c_{kA\sigma}^{\dag}c_{kA\sigma}-c_{kB\sigma}^{\dag}c_{kB\sigma}], \label{eq4}
\end{eqnarray}
where we have defined $f(k)=e^{-ik_{x}}+2e^{ik_{x}/2}\cos(\frac{\sqrt{3}}{2}k_{y})$, $\gamma(k)=\sin(\sqrt{3}k_{y})-2\cos(\frac{3}{2}k_{x})\sin(\frac{\sqrt{3}}{2}k_{y})$ and $A$, $B$ representing two nonequivalent sublattices of the honeycomb lattice, respectively. Then, by diagonalizing the above Hamiltonian, one
obtains the quasiparticle energy band as
\begin{eqnarray}
E_{k\sigma\pm}=\pm\sqrt{t^{2}|f(k)|^{2}+4t'^{2}\gamma(k)^{2}}, \label{eq5}
\end{eqnarray}
which preserves the particle-hole symmetry and also the spin degeneracy. It is well-known that for $3\sqrt{3}t'<t$, the excitation gap mainly opens near six Dirac points (Only two of them are nonequivalent in fact).\cite{Rachel} Then, expanding both $f(k)$ and $\gamma(k)$ near two nonequivalent Dirac points $\pm\vec{K}=\pm(0,\frac{4\pi}{3\sqrt{3}})$, respectively, the gap can be found as $\Delta_{gap}=6\sqrt{3}t'$ and the quasiparticle energy reads $E_{q\sigma\pm}\simeq\pm\sqrt{(\frac{3}{2}tq)^{2}+(3\sqrt{3}t')^{2}}$ with $q=(q_{x},q_{y})\equiv(k_{x},k_{y}\mp\frac{4\pi}{3\sqrt{3}})$.

The most interesting property of the Kane-Mele model is that it gives rise to a vanished charge Hall conductance but a nonzero quantized spin Hall conductance in contrast to the usual quantum Hall or quantum anomalous Hall  effect.\cite{DSheng2006} The corresponding quantized spin Hall conductance can be readily calculated by $\sigma_{SH}=(C_{1}^{\uparrow}-C_{1}^{\downarrow})\frac{1}{2e}\frac{e^{2}}{h}$ where the (spin) Chern number $C_{1}^{\sigma=\uparrow/\downarrow}$ for each spin part is defined by $C_{1}^{\sigma}=\frac{1}{4\pi}\int dk_{x}dk_{y}\hat{\textbf{d}_{\sigma}}\cdot(\frac{\partial\hat{\textbf{d}_{\sigma}}}{\partial k_{x}}\times \frac{\partial\hat{\textbf{d}_{\sigma}}}{\partial k_{y}})$.\cite{DSheng2006,Prodan2009,Qi2011} For the Kane-Mele model, the single particle Hamiltonian can be rewritten as $\hat{h}_{k\sigma}=\textbf{d}_{\sigma}(k)\cdot \hat{\bf{\sigma}}$ with $\textbf{d}_{\sigma}=(-t \text{Re} f(k),t \text{Im} f(k),2t'\sigma\gamma(k))$, $\hat{\textbf{d}_{\sigma}}=\textbf{d}_{\sigma}/|\textbf{d}_{\sigma}|$ and $\hat{\bf{\sigma}}$ being the usual Pauli matrices. Then, by inserting the expression of $\hat{\textbf{d}_{\sigma}}$ into the formula of $C_{1}^{\sigma}$, one obtains $C_{1}^{\uparrow}=-C_{1}^{\downarrow}=1$ and $\sigma_{SH}=\frac{e}{h}$ while the charge Hall conductance $\sigma_{H}=(C_{1}^{\uparrow}+C_{1}^{\downarrow})\frac{e^{2}}{h}=0$.\cite{DSheng2006,Qi2011} In literature, the ground-state of the Kane-Mele model is often called the quantum spin Hall insulator (topological insulator) since it
exhibits a quantized spin Hall conductance, which has the gapped bulk spectrum with a gapless helical edge-state protected by the time-reversal symmetry.\cite{Kane2006}

\subsection{The low-energy effective theory for the Kane-Mele model}
For the discussion of the low-energy physics, an effective $2+1$D massive Dirac action can be obtained by expanding original Kane-Mele model around two nonequivalent Dirac points $\pm\vec{K}$,
\begin{eqnarray}
S_{\text{K}}=\int d^{2}xd\tau \mathcal{L}_{0}=\int d^{2}xd\tau\sum_{a\sigma}[\bar{\psi}_{a\sigma}(\gamma_{\mu}\partial_{\mu}+m\sigma)\psi_{a\sigma}], \nonumber
\end{eqnarray}
where $\gamma_{\mu}=(\tau_{z},\tau_{x},\tau_{y})$ and $\partial_{\mu}=(\partial_{\tau},\partial_{x},\partial_{y})$
with $\tau_{z},\tau_{x},\tau_{y}$ the usual Pauli matrices. Here the same indices mean summation. We introduce the effective mass $m=-3\sqrt{3}t'$ of Dirac fermions and set the effective Fermi velocity $v_{F}=\frac{3}{2}t$ to unit. The Dirac fields are defined as $\psi_{1\sigma}=(c_{1A\sigma},c_{1B\sigma})^{T}$, $\psi_{2\sigma}=(c_{2A\sigma},-c_{2B\sigma})^{T}$ and $\bar{\psi}_{a\sigma}=\psi^{\dag}_{a\sigma}\gamma_{0}$ with $a=1,2$ denoting the states near the two nonequivalent Dirac points $\pm\vec{K}=\pm(0,\frac{4\pi}{3\sqrt{3}})$ and $T$ implying the transposition manipulation. We should remind the reader that the spin-up and spin-down fields $\psi_{a\sigma}$ acquire opposite mass term as can be seen from the above effective action, which leads to vanished charge Hall conductance but a nonzero quantized spin Hall conductance.

Having obtained the effective massive Dirac action in 2+1D, it is interesting to see its physical response to the external electromagnetic field $A_{\mu}^{c}=(i\phi^{c},A_{x}^{c},A_{y}^{c})$ with $\phi^{c}$ and $\vec{A}^{c}=(A_{x}^{c},A_{y}^{c})$ representing
the usual scalar and vector potentials, respectively. The electromagnetic field can be readily introduced into the Dirac action by the conventional minimal coupling, namely, $\partial_{\mu}\rightarrow\partial_{\mu}-iA_{\mu}^{c}$.
(The charge $e$ is setting to unit for simplicity.) However, one will find that the Chern-Simons action for the electromagnetic field $A_{\mu}^{c}$ vanishes due to the opposite mass for different spin-flavors, which means the charge Hall conductance is zero. (This is consistent with the results of the previous subsection.) But, one can add the artificial external spin gauge-field $A_{\mu}^{s}$ to see the nontrivial spin response.\cite{Grover2008} Thus, the resulting effective Dirac action coupled with the external electromagnetic and spin fields reads
\begin{equation}
S=\int d^{2}xd\tau\sum_{a\sigma}\bar{\psi}_{a\sigma}[\gamma_{\mu}(\partial_{\mu}-iA_{\mu}^{c}-\frac{1}{2}i\sigma A_{\mu}^{s})+m\sigma]\psi_{a\sigma}.\label{eq6}
\end{equation}

By integrating out the Dirac fields, we get an effective mutual Chern-Simons action, which represents the electromagnetic response of the massive Dirac fermions to the external electromagnetic field $A_{\mu}^{c}$ and artificial
external spin gauge-field $A_{\mu}^{s}$ (for details, one can refer to Appendix A),
\begin{equation}
S_{\text{CS}}=2 \int d^{2}xd\tau\left[\frac{-im\sigma}{8\pi|m\sigma|}\epsilon^{\mu\nu\lambda}(A_{\mu}^{c}+\frac{\sigma}{2}A_{\mu}^{s})\partial_{\nu}(A_{\lambda}^{c}+\frac{\sigma}{2}A_{\lambda}^{s})\right],\label{eq7}
\end{equation}
where the pre-factor 2 comes from the nonequivalent Dirac points, $\epsilon^{\mu\nu\lambda}$ is the usual all-antisymmetric tensor and we have dropped out the regular Maxwell term ($\sim F^{2}_{\mu\nu}$) since the low energy physics is dominated by the Chern-Simons term alone. We should emphasize that although the effective Chern-Simons action is used here, it does not imply any fractionalization or nontrivial topological order (A characteristic signature of the topological order is the ground-state degeneracy depending on the topology of the system.) because no emergent gauge fields or fractionalized quasiparticles exist in the present case.\cite{Wen2004}

Since the effective mass $m=-3\sqrt{3}t'<0$, the above Chern-Simons action can be simplified to the following form
\begin{equation}
S_{\text{CS}}=\int d^{2}xd\tau\left[\frac{i}{4\pi}\epsilon^{\mu\nu\lambda}(A_{\mu}^{c}\partial_{\nu}A_{\lambda}^{s}+A_{\mu}^{s}\partial_{\nu}A_{\lambda}^{c})\right].\label{eq8}
\end{equation}

Now, one can see that the quantized spin Hall conductance $\sigma_{SH}=\frac{1}{2\pi}$ from $J_{x}^{s}=\frac{\partial S_{CS}}{\partial A_{x}^{s}}|_{\vec{A}^{s/c}\rightarrow0}=\frac{2i}{4\pi}(\partial_{y}A_{0}-\partial_{0}A_{y})=\frac{2}{4\pi}E_{y}$. If we reintroduce $h=2\pi\hbar$ and charge $e$, this spin hall conductance reads $\sigma_{SH}=\frac{e}{h}$, which reproduces the correct result in terms of spin Chern number in the previous subsection.
We note that the existence of the spin Hall conductance relies on the existence of the charge degree of freedom since the spin gauge-field $A_{\mu}^{s}$ couples to the electromagnetic field $A_{\mu}^{c}$ in the Chern-Simons action. Otherwise, if the elementary particles (here the electrons) do not carry charge, their spin Hall conductance is obviously zero. The case here is rather different to the spin quantum Hall effect,\cite{Senthil1999,Lu2012c} whose
spin Hall conductance does not rely on the charge degree of freedom.

\section{The antiferromagnetic spin-density-wave state} \label{sec3}
In this section, we will present the full discussion of the antiferromagnetic spin-density-wave state including the mean-field equations, the effective action and the edge-states.

\subsection{The antiferromagnetic spin-density-wave state}
For the case with $J_{\parallel}\gg J_{\perp}$, in general, one expects that the antiferromagnetic SDW state to be the stable ground-state of Kondo lattice model on the honeycomb lattice due to its bipartite feature.\cite{Tsunetsugu} To study the possible antiferromagnetic ordered state, we assume that no Kondo screening exists ($V=0$) while magnetic order parameters $m_{c},m_{d}$ are not vanishing in the mean-field Hamiltonian
Eq. (\ref{eq2}). Then, the resultant Hamiltonian reads
\begin{eqnarray}
&&H_{\text{MF}}=H_{\text{c}}+H_{\text{d}}+E_{0},\nonumber\\
&&H_{\text{c}}=\sum_{k\sigma}-t[f(k)c_{kA\sigma}^{\dag}c_{kB\sigma}+f^{\star}(k)c_{kB\sigma}^{\dag}c_{kA\sigma}] \nonumber\\
&& + (2t'\sigma\gamma(k)+\frac{J_{\parallel}}{2}\sigma m_{d})[c_{kA\sigma}^{\dag}c_{kA\sigma}-c_{kB\sigma}^{\dag}c_{kB\sigma}],\nonumber\\
&&H_{\text{d}}=\frac{J_{\parallel}}{2}\sum_{k\sigma}[\sigma(-m_{c}d_{kA\sigma}^{\dag}d_{kA\sigma}+-m_{c}d_{kB\sigma}^{\dag}d_{kB\sigma}),\nonumber\\
&&E_{0}=N_{s}2J_{\parallel}m_{d}m_{c}. \nonumber
\end{eqnarray}

\subsubsection{Possibility of a SDW state with a quantized spin Hall conductance}
Here we see that the conduction electrons $c$ and local pseudofermions $d$ are clearly decoupled at the present mean-field level. The nonzero magnetic order parameters $m_{c},m_{d}$ mean that the state obtained by such a hamiltonian is magnetically ordered.
Besides, we may note that the conduction electron part $H_{\text{c}}$ is similar to the Kane-Mele model Eq. (\ref{eq3}) with an extra effective coupling from the magnetic order of local spins.
Based on such an observation, if the conduction electron part is able to support a quantized spin Hall conductance, the obtained antiferromagnetic state should be identified as the topological one.
Using the same method in the discussion of Kane-Mele model in the previous section, the spin Chern numbers are calculated as $C_{1}^{\uparrow}=-C_{1}^{\downarrow}=\Theta(2t'3\sqrt{3}-\frac{J_{\parallel}}{2}m_{d})$, where we assume $t',m_{d},J_{\parallel}>0$ and the step function $\Theta(x)$ is defined by $\Theta(x)=1$ ($=0$) for $x>0$ ($x<0$). Therefore, if $C_{1}^{\uparrow},C_{1}^{\downarrow}$ are nonzero, we will
have an antiferromagnetic state with a quantized spin Hall conductance ($\sigma_{SH}=(C_{1}^{\uparrow}-C_{1}^{\downarrow})\frac{1}{2e}\frac{e^{2}}{h}$). To find such interesting possibility, we have to solve the above mean-field Hamiltonian below (Note that $m_{d}$ has to be solved by the mean-field equations).

\subsubsection{The mean-field equations of the SDW states}
Diagonalizing the mean-field Hamiltonian, we can easily derive ground-state energy of the antiferromagnetic SDW state per site as
\begin{eqnarray}
E_{g}^{AFM}&&=J_{\parallel}m_{c}(2m_{d}-1) \nonumber\\
&&-\frac{1}{N_{s}}\sum_{k\sigma}\sqrt{(\frac{J_{\parallel}m_{d}}{2}+\gamma(k))^{2}+t^{2}|f(k)|^{2}}\nonumber
\end{eqnarray}
and two self-consistent equations from minimizing $E_{g}^{AFM}$ with respect to magnetization $m_{d}$ and $m_{c}$, respectively.
\begin{eqnarray}
&&J_{\parallel}m_{c}(2m_{d}-1)=0,\nonumber\\
&&m_{c}=\frac{2}{4N_{s}}\sum_{k}\frac{J_{\parallel}m_{d}/2+\gamma(k)}{\sqrt{(\frac{J_{\parallel}m_{d}}{2}+\gamma(k))^{2}+t^{2}|f(k)|^{2}}}.\nonumber
\end{eqnarray}

For further analytical treatment, one may use a simplified linear density of state (DOS) $\rho(\varepsilon)=|\varepsilon|/\Lambda^{2}$ when transforming the summation over momentum $k$ into integral on energy $\varepsilon$ with $\Lambda\simeq2.33t$ being high-energy cutoff. Thus, $t|f(k)|$ is replaced by $|\varepsilon|$ to simplify corresponding calculations and $\gamma(k)$ is replaced by $\pm3\sqrt{3}t'$ near two nonequivalent Dirac points $\pm\vec{K}=\pm(0,\frac{4\pi}{3\sqrt{3}})$.

From these two equations, one obtains $m_{d}=1/2$ and
\begin{equation}
m_{c}=\frac{1}{2\Lambda^{2}} \sum_{\sigma}R_{\sigma}(\sqrt{\Lambda^{2}+R^{2}_{\sigma}}-\sqrt{R_{\sigma}^{2}})\nonumber
\end{equation}
with $R_{\sigma}=3\sqrt{3}\sigma t'+J_{\parallel}/4$
while the ground-state energy per site for the antiferromagnetic SDW state reads
\begin{equation}
E_{g}^{AFM}=-\frac{2}{3\Lambda^{2}}\sum_{\sigma}[(\Lambda^{2}+R^{2}_{\sigma})^{3/2}-(R_{\sigma}^2)^{3/2}].\label{eq9}
\end{equation}

From above equations, we observe that the local spins are fully polarized ($m_{d}=1/2$) while the conduction electrons have small magnetization. It is also noted that when $t'=0$ (no next-nearest-neighbor hopping), the above $m_{c}$ correctly recovers the value in our previous work.\cite{Zhong2012b} Meanwhile, the low-lying quasiparticle excitations in the antiferromagnetic SDW state have the energy $E_{\pm\pm\sigma}(k)=\pm\sqrt{(3tk/2)^{2}+( J_{\parallel}/4\pm3\sqrt{3}t')^{2}}$ and $E_{\pm\pm\sigma}(k)=\pm J_{\parallel}m_{c}/2$. The former corresponds to the conduction electrons while the latter, which is nondispersive, relates to the local spins/pseudofermions.  It should be noted that the gap around the Dirac points only closes when the condition $J_{\parallel}/4=3\sqrt{3}t'$ is fully satisfied. Otherwise, any low-lying quasiparticle excitations in the antiferromagnetic SDW state are clearly gapped. Thus, we may conclude that the antiferromagnetic SDW state we obtained is mainly an insulating state (except for the case with $J_{\parallel}/4=3\sqrt{3}t'$) with fully polarized local spins ($m_{d}=1/2$) while conduction electrons only partially polarize ($m_{c}<1/2$). This feature is similar to the previous study on square lattice, thus confirms the validity of our current treatment.\cite{Zhang2000}

Furthermore, since we have calculated the order parameter $m_{d}=1/2$, according to the discussion in the previous paragraph, the spin Chern numbers are found as $C_{1}^{\uparrow}=-C_{1}^{\downarrow}=\Theta(2t'3\sqrt{3}-\frac{J_{\parallel}}{4})$. Therefore, when $2t'3\sqrt{3}>\frac{J_{\parallel}}{4}$, we have $C_{1}^{\uparrow}=-C_{1}^{\downarrow}=1$ and the
spin Hall conductance is quantized to $\sigma_{SH}=(C_{1}^{\uparrow}-C_{1}^{\downarrow})\frac{1}{2e}\frac{e^{2}}{h}=\frac{e}{h}$, which has the identical value to the Kane-Mele model.
Considering that this state also has antiferromagnetic SDW long-ranged ordered state, we may call it topological SDW (T-SDW) state. In contrast, when $2t'3\sqrt{3}<\frac{J_{\parallel}}{4}$, the spin
Chern numbers vanish and we only have a normal SDW (N-SDW) state. There also exists a particular point where $2t'3\sqrt{3}=\frac{J_{\parallel}}{4}$. This case corresponds to
the vanished gap for quasiparticle excitations and can be identified as a topological quantum phase transition between the N-SDW state and the T-SDW one. More details will be pursued in the next subsection.

\subsection{The physics of the T-SDW state}
Having established the existence of the T-SDW and N-SDW states, we will use effective theories to get deeper insight into those states and their corresponding topological quantum phase transition in this and
next two subsections.

\subsubsection{The T-SDW state}
Performing the same treatment as for the Kane-Mele model in Sec.\ref{sec2} on the mean-field Hamiltonian,  we obtains the following effective action for the antiferromagnetic SDW state
\begin{eqnarray}
&&S=\int d^{2}xd\tau\sum_{a\sigma}\bar{\psi}_{a\sigma}\left[\gamma_{\mu}(\partial_{\mu}-iA_{\mu}^{c}-i\frac{\sigma}{2}A_{\mu}^{s})+m_{a\sigma}\right]\psi_{a\sigma},\nonumber\\
\label{eq10}
\end{eqnarray}
where the effective mass is defined as $m_{1\uparrow}=m-J_{\parallel}/4$, $m_{2\downarrow}=-m-J_{\parallel}/4$, $m_{1\downarrow}=-m+J_{\parallel}/4$ and $m_{2\uparrow}=m+J_{\parallel}/4$ with $m=-3\sqrt{3}t'$. [Note that
$m_{1\uparrow}<0,m_{1\downarrow}>0$.] Then, it is straightforward to derive an effective Chern-Simons action by integrating out the Dirac fermions
\begin{eqnarray}
S_{\text{CS}}&&=\int d^{2}xd\tau\left[\sum_{a\sigma}\frac{m_{a\sigma}}{|m_{a\sigma}|}\frac{-i}{8\pi}\epsilon^{\mu\nu\lambda}(A_{\mu}^{c}+\frac{\sigma}{2}A_{\mu}^{s})\partial_{\nu}(A_{\lambda}^{c}+\frac{\sigma}{2}A_{\lambda}^{s})\right]\nonumber\\
&&=\int d^{2}xd\tau\left[\left(\sum_{a\sigma}\frac{m_{a\sigma}\sigma}{|m_{a\sigma}|}\right)\frac{-i}{8\pi}\epsilon^{\mu\nu\lambda}A_{\mu}^{c}\partial_{\nu}A_{\lambda}^{s}\right]\nonumber\\
&&=\int d^{2}xd\tau\left[(2\text{sgn}(m+J_{\parallel}/4)-2)\frac{-i}{8\pi}\epsilon^{\mu\nu\lambda}A_{\mu}^{c}\partial_{\nu}A_{\lambda}^{s}\right],\nonumber
\end{eqnarray}
where the sign function $\text{sgn}(x)=1$ for $x>0$ and $\text{sgn}(x)=-1$ for $x<0$ and we have used the fact $m_{1\uparrow}<0,m_{1\downarrow}>0$.
It is easy to see that a quantized spin Hall conductance with the value $\sigma_{SH}=e^{2}/h$ is obtained if $-m=3\sqrt{3}t'>J_{\parallel}/4$ (See the second subsection of Sec.\ref{sec2} for the calculation of the quantized spin Hall conductance.). In contrast, when $3\sqrt{3}t'$ is smaller than $J_{\parallel}/4$, no such quantized spin Hall conductance can be found and the corresponding effective Chern-Simon term vanishes.

Therefore, it seems that even in the antiferromagnetic SDW state, there exists a quantized spin Hall conductance if the quasiparticle gap is still dominated by the next-nearest neighbor hopping ($3\sqrt{3}t'$) rather than the antiferromagnetic order ($J_{\parallel}/4$). Thus, we have uncovered a topological antiferromagnetic SDW state, namely, the T-SDW state with a quantum spin Hall effect for $3\sqrt{3}t'>J_{\parallel}/4$.

\subsubsection{The hydrodynamics approach and the topological property of the T-SDW state}
It is also interesting to see the topological properties in the T-SDW state. Since the T-SDW state has nonzero spin Chern numbers or quantized spin Hall conductance,
follow the hydrodynamics approach for integer/fractional quantum Hall (IQH/FQH) states,\cite{Wen2004} we first write down the
the conservation current for each spin-flavor $J_{1\mu}=\frac{1}{2\pi}\epsilon^{\mu\nu\lambda}\partial_{\nu}a_{1\lambda},J_{2\mu}=\frac{1}{2\pi}\epsilon^{\mu\nu\lambda}\partial_{\nu}a_{1\lambda}$
(here 1 for spin-up and 2 for spin-down electrons)

Due to the fact that the spin-up (down) electrons has spin Chern number $C_{1}^{\uparrow}=1$ ($C_{1}^{\downarrow}=-1$), we can attribute Chern-Simons terms to each flavor as
$\frac{-i}{4\pi}\epsilon^{\mu\nu\lambda}a_{1\mu}\partial_{\nu}a_{1\lambda}$ for spin-up and $\frac{+i}{4\pi}\epsilon^{\mu\nu\lambda}a_{2\mu}\partial_{\nu}a_{2\lambda}$ for spin down ones.

Next, the conserved currents are coupled to the external electromagnetic field $A_{\mu}^{c}$ with identical charge ($iA_{\mu}^{c}(J_{1\mu}-J_{2\mu})$) and coupled to the spin gauge-field $A_{\mu}^{s}$ with charge $1/2$ (spin-up electrons) and $-1/2$ (spin-down electrons) ($\frac{i}{2}A_{\mu}^{s}(J_{1\mu}-J_{2\mu})$). Combining these terms, we have the following effective abelian Chern-Simons action
\begin{eqnarray}
S&&=\int d^{2}xd\tau[K_{IJ}\frac{-i}{4\pi}\epsilon^{\mu\nu\lambda}a_{I\mu}\partial_{\nu}a_{J\lambda}+\frac{i}{2\pi}\epsilon^{\mu\nu\lambda}q_{I}A_{\mu}^{c}\partial_{\nu}a_{I\lambda}\nonumber\\
&&+\frac{i}{2\pi}\epsilon^{\mu\nu\lambda}\frac{s_{I}}{2}A_{\mu}^{s}\partial_{\nu}a_{I\lambda}],\label{eq11}
\end{eqnarray}
where we have defined the so-called $K$ matrix as $K_{11}=1,K_{22}=-1,K_{12}=K_{21}=0$ for $I,J=1,2$  and the corresponding charge and spin vectors read $q=(1,1)^{T}$ and $s=(1,-1)^{T}$, respectively.\cite{Wen2004} A care reader may note that we have introduced the spin gauge-field $A_{\mu}^{s}$ and the spin vector $s_{I}$, which do not appear in usual Chern-Simons actions for description of topological states. For the physical observable, the quantized charge Hall conductance is calculated as $\sigma_{H}=\nu\frac{e^{2}}{h}=0$ with $\nu=q^{T}K^{-1}q=0$ and the quantized spin Hall conductance reads $\sigma_{SH}=\nu_{s}\frac{e}{2h}=\frac{e}{h}$ with $\nu_{s}=s^{T}K^{-1}q=2$, thus the effective action correctly reproduces the quantized spin Hall conductance and vanished charge Hall conductance obtained in the previous subsection.

Additionally, some readers may wonder whether there exists a nontrivial topological order since the effective theory Eq. (\ref{eq11}) is a Chern-Simons action. This issue can be seen as follows. Firstly, one finds $Det[K]=1$, which means no state with topological order and fractional excitations are involved in the present system \cite{Lu2012}. In contrast, systems supporting fractionalized excitations should require a $K$-matrix with $|Det[K]|>1$ since the ground-state degeneracy on a torus is equal to $Det[K]$.\cite{Senthil2012}). Secondly, if one examines the statistics of the elementary quasiparticle (the electron) in terms of the so-call exchange statistical angle, without any surprise, one will find such angle is $\pi$ which implies that the exchange of two identical quasiparticle gives rise to a $\pi$ phase in their wavefunction. Therefore, we conclude that the elementary quasiparticle is the usual electron as expected.

\subsubsection{The edge-state of the T-SDW state}
It is well-known that many topological states, e.g. QSH, IQH and FQH insulators, have nontrivial edge-states in contrast to trivial band insulators.\cite{Qi2011,Wen2004} Therefore, we will discuss property of the edge-state for the T-SDW state.

\begin{figure}
\includegraphics[width=0.8\columnwidth]{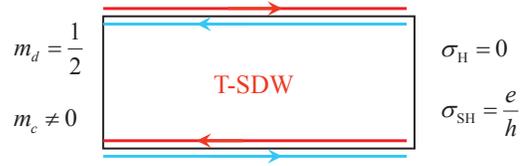}
\caption{\label{fig:1} The basic properties of the topological antiferromagnetic SDW (T-SDW) state is that it has a quantum spin Hall conductance $\sigma_{SH}=e/h$, a zero charge hall conductance $\sigma_{H}=0$
and antiferromagnetic orders (The magnetic moment of local spins is $m_{d}=1/2$ while the conduction electrons has $m_{c}\neq0$).
The edge-state is the helical state which has counter-moving modes locked to the spin direction. The red mode denotes the spin-up electrons while the blue one represents the spin-down electrons.}
\end{figure}

Since the topological features are described in terms of the Chern-Simons action Eq. (\ref{eq11}), according to the standard bulk-edge correspondence for the effective abelian Chern-Simons theory, the gapless edge states is
described by two counter-moving bosonic modes\cite{Wen2004}
\begin{eqnarray}
S_{\text{edge}}&&=\int dxd\tau\frac{1}{4\pi}[-i\partial_{\tau}\phi_{1}\partial_{x}\phi_{1}+c\partial_{x}\phi_{1}\partial_{x}\phi_{1}\nonumber\\
&&+i\partial_{\tau}\phi_{2}\partial_{x}\phi_{2}+c\partial_{x}\phi_{2}\partial_{x}\phi_{2}]\label{eq12}
\end{eqnarray}
with $c$ denoting the non-universal velocity of edge states and $\phi_{I}$ being the bosonic representation for the two edge-state modes ($I=1$ for the spin-up mode while $I=2$ for the spin-down mode).
Because the right (left) mover carries spin-up (spin-down) degree of freedom, this edge-state is just the helical edge-state which is widely studied in the field of the quantum spin Hall effect/two-dimensional topological insulator.\cite{Qi2011} To see this correspondence more clearly, one can refermionize above action by introducing the fermion operator $\psi_{I}\propto e^{i\phi_{I}}$ and the resulting action reads
\begin{eqnarray}
S_{\text{edge}}=\int dx d\tau[\psi^{\dag}_{1}(\partial_{\tau}-ic\partial_{x})\psi_{1}+\psi^{\dag}_{2}(\partial_{\tau}+ic\partial_{x})\psi_{2}].\nonumber
\end{eqnarray}
And the corresponding Hamiltonian formalism is simple
\begin{eqnarray}
H_{\text{edge}}=\int dx[\psi^{\dag}_{1}(-ic\partial_{x})\psi_{1}+\psi^{\dag}_{2}(ic\partial_{x})\psi_{2}],\label{eq13}
\end{eqnarray}
which is just the usual helical edge-state in QSH insulator.\cite{Qi2011} [The chiral bosonic field $\phi_{I}$ is now rewritten as $\phi_{1}=\phi-\theta,\phi_{1}=-\phi-\theta$ where $\psi_{1}\propto e^{i(\phi-\theta)},\psi_{2}\propto e^{i(-\phi-\theta)}$ and $\phi,\theta$ satisfy the standard bosonized Hamiltonian $
H_{\text{edge}}=\frac{c}{2\pi}\int dx[(\partial_{x}\theta)^{2}+(\partial_{x}\phi)^{2}]$.] The basic properties of the topological antiferromagnetic SDW (T-SDW) state is summarized in Fig.~\ref{fig:1}.

\subsubsection{The stability of edge-state of the T-SDW state and conservation of the time-reversal symmetry at edges}
At the free-particle level, i.e. Eq. (\ref{eq12}) or (\ref{eq13}), there does not exist an excitation gap for bosonic modes or electrons at edge.
However, when interaction or impurity effects are involved, the stability of the gaplessness of the above edge-state should be inspected.

Firstly, if the conservation of particle number is preserved, in general, the local interacting terms will be the following form\cite{Giamarchi}
\begin{eqnarray}
&&H_{\text{int}}=H_{1}+H_{2}+H_{3}\nonumber\\
&&H_{1}=g_{1}\int dx[\psi^{\dag}_{1}\psi_{1}\psi^{\dag}_{1}\psi_{1}+\psi^{\dag}_{2}\psi_{2}\psi^{\dag}_{2}\psi_{2}]\nonumber\\
&&H_{2}=g_{2}\int dx[\psi^{\dag}_{1}\psi_{1}\psi^{\dag}_{2}\psi_{2}]\nonumber\\
&&H_{3}=g_{3}\int dx[\psi^{\dag}_{1}\psi^{\dag}_{1}\psi_{2}\psi_{2}+\psi^{\dag}_{2}\psi^{\dag}_{2}\psi_{1}\psi_{1}].\nonumber
\end{eqnarray}
Obviously, the dangerous umklapp term $H_{3}$ vanishes due to $\psi_{I}\psi_{I}=\psi^{\dag}_{I}\psi^{\dag}_{I}=0 (I = 1,2)$. $H_{1},H_{2}$ do not vanish but they cannot gap out the edge states since they only correspond to the usual forward scattering.

Secondly, besides these interacting terms, the mass term $H_{m}=M\int dx[\psi^{\dag}_{1}\psi_{2}+\psi^{\dag}_{2}\psi_{1}]$, which can result from weak impurity scattering, is important.\cite{Giamarchi}
Adding this backscattering term to the free Hamiltonian part Eq. (\ref{eq13}), one can see an excitation gap $\Delta=2|M|$. Thus, if no symmetry is imposed, this backscattering term will destroy the gaplessness
of the edge-state of T-SDW state described by Eq. (\ref{eq13}).

We remind the reader that in the field of QSH insulator, one assumes the time-reversal symmetry is still preserved at the edge, so such impurity scattering
term is prohibited since it is not invariant under the time-reversal transformation ($\psi_{1}\rightarrow\psi_{2},\psi_{2}\rightarrow-\psi_{1},i\rightarrow-i$).
In our case, the bulk has the antiferromagnetic SDW order and this magnetic order inevitably breaks the time-reversal symmetry in the bulk.
As for the edge, we suspect that the time-reversal invariance is in fact persevered in the T-SDW state although it is broken in the bulk. This conjecture is motivated by the numerical study of the Bernevig-Hughes-Zhang model with Hubbard-$U$ term.\cite{Yoshida2013b} In their case, the T-SDW state is also found with nontrivial quantized spin Hall conductance, whose value is found to approach unit ($\frac{e}{h}$). Based on this observation, those authors deduce that the antiferromagnetic order is suppressed at the edges, and the helical edge states are topologically protected against magnetic instability, in spite of the existence of the bulk antiferromagnetic order.\cite{Yoshida2013b}
Therefore, we will assume that the gaplessness of helical edge-state Eq. (\ref{eq13}) of T-SDW state is stable to impurity as time-reversal symmetry is conserved at the edge.

Since the spin-up and spin-down part of the gapless edge-state contributes $\frac{e}{2h}$ to the spin Hall conductance,\cite{DSheng2006} one obtains that the total spin Hall conductance is quantized to $\frac{e}{h}$, as expected. Obviously, this robust quantized value for the spin Hall conductance is protected by the time-reversal symmetry as we have assumed. However, we should emphasize that the nonzero quantized spin Chern number does not necessarily
imply the existence of the gapless edge-state, as shown in Ref.\onlinecite{LSheng2011}, where the authors found that the spin Chern number is well quantized but the edge-states are gapped when the time-reversal symmetry of the Kane-Mele model is broken by the Rashba spin-orbit coupling or the magnetic exchange field. Therefore, one should be careful when naively using the bulk-edge correspondence.

\subsection{The N-SDW state}
When $J_{\parallel}/4>3\sqrt{3}t'$, the Chern-Simons term vanishes and this indicates that this case is just the usual antiferromagnetic SDW state, namely, the N-SDW. If we examine the T-SDW and N-SDW from the point of view of symmetry, we find that these two SDW states have identical physical symmetry with broken spin-rotation invariance. However, from the discussion of previous sections, the spin response of them is rather different (The former has quantized spin Hall conductance while the latter one does not.) which can be considered as
a useful criteria to distinguish those two states. In this sense, the T-SDW state provides a simple example beyond the symmetry-breaking based classification of states of matter.\cite{Wen2004}

\subsection{The topological quantum phase transition}
When $J_{\parallel}/4=3\sqrt{3}t'$, the effective mass of two Dirac fields are zero $m_{2\uparrow}=m_{2\downarrow}=0$ while others are still massive $m_{1\uparrow}=-m_{1\downarrow}=-J_{\parallel}/2<0$.
Since $J_{\parallel}/4<3\sqrt{3}t'$ and $J_{\parallel}/4>3\sqrt{3}t'$ correspond to T- and N-SDW states, which have rather different topological properties (the T-SDW state has quantized spin Hall conductance while the N-SDW does not.),
one may consider the point $J_{\parallel}/4=3\sqrt{3}t'$ as a topological quantum phase transition point. To sharpen our understanding on this topological quantum phase transition point, one could integrate out the massive
Dirac field and get the following effective action
\begin{eqnarray}
&&S=S_{\text{CS}}+S_{2\uparrow}+S_{2\downarrow},\nonumber\\
&&S_{\text{CS}}=\int d^{2}xd\tau[\frac{i}{4\pi}\epsilon^{\mu\nu\lambda}A_{\mu}^{c}\partial_{\nu}A_{\lambda}^{s}].\nonumber\\
&&S_{2\uparrow}=\int d^{2}xd\tau[\bar{\psi}_{2\uparrow}\gamma_{\mu}(\partial_{\mu}-iA_{\mu}^{c}-\frac{i}{2}A_{\mu}^{s})\psi_{2\uparrow}],\nonumber\\
&&S_{2\downarrow}=\int d^{2}xd\tau[\bar{\psi}_{2\downarrow}\gamma_{\mu}(\partial_{\mu}-iA_{\mu}^{c}+\frac{i}{2}A_{\mu}^{s})\psi_{2\downarrow}],\nonumber
\end{eqnarray}
where we have integrated out massive $\psi_{1\uparrow}$ and $\psi_{1\downarrow}$. Clearly, this quantum phase transition point is described by two gapless
free Dirac fields and its thermaldynamic and transport properties are familiar. According to the conventional scaling theory,\cite{Continentino} the thermaldynamic properties are described by the free energy
$f(T)\sim T^{\frac{d+z}{z}}$ where $T,d,z$ denoting temperature, spatial dimension and dynamical critical exponent. For our present system, $d=2$ and $z=1$, which reflects the emergent Lorentz symmetry (Time and space have the same scaling under the standard scaling transformation.). Therefore, the specific heat is found as $C_{v}\sim T\frac{\partial^{2}f(T)}{\partial^{2}T}\sim T^{2}$. The magnetization has the similar behavior as the specific heat
$M(T)\sim T^{2}$ due to the same scaling of the external magnetic field and the temperature. Thus, the magnetic susceptibility reads $\chi(T)\sim \frac{\partial M}{\partial B}|_{B\rightarrow T}\sim T$.

For the electromagnetic response, we can integrate out $\psi_{2\uparrow}$ and $\psi_{2\downarrow}$ to get the following action
\begin{eqnarray}
S_{eff}=\int\frac{d^{3}q}{(2\pi)^{3}}\left(2A_{\mu}^{c}\Pi_{\mu\nu}(q)A_{\nu}^{c}+\frac{2}{4e^{2}}A_{\mu}^{s}\Pi_{\mu\nu}(q)A_{\nu}^{s}\right),\nonumber
\end{eqnarray}
where the polarization function $\Pi_{\mu\nu}(q)=\frac{e^{2}}{16}(\delta_{\mu\nu}-\frac{q_{\mu}q_{\nu}}{q^{2}})\sqrt{q^{2}}$ is calculated in Appendix \ref{B}.
Therefore, the charge conductance is found as $\sigma_{xx}(\omega)=\sigma_{yy}(\omega)=\frac{2}{-i\omega}\Pi_{\mu\nu}(\vec{q},i\omega_{n}\rightarrow\omega)|_{\vec{q}\rightarrow0}=\frac{2}{-i\omega}\frac{e^{2}}{16}\sqrt{\vec{q}^{2}-\omega^{2}}|_{\vec{q}\rightarrow0}=\frac{e^{2}}{8}=\frac{\pi e^{2}}{8\pi\hbar}=\frac{\pi}{4}\frac{e^{2}}{\hbar}$ when $\sigma_{xy}(\omega)=\sigma_{yx}(\omega)=0$. (Note that the charge conductance is independent of the frequency.) As comparison, we note that the charge conductance are all vanished in both T and N-SDW states.

\begin{figure}
\includegraphics[width=0.8\columnwidth]{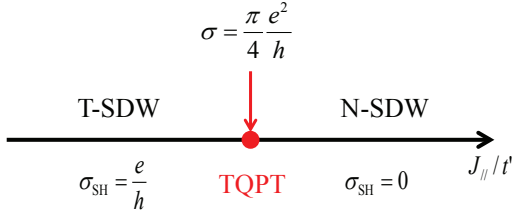}
\caption{\label{fig:2} The topological quantum phase transition (TQPT) between the normal antiferromagnetic SDW (N-SDW) state and the topological antiferromagnetic SDW (T-SDW) state with a quantum spin Hall conductance $\sigma_{SH}=e^{2}/h$. The TQPT is located at $3\sqrt{3}t'=J_{\parallel}/4$ with $\Lambda\simeq2.33t$ being high-energy cutoff. As noted, there exists a nonzero charge conductance $\sigma=\frac{\pi}{4}\frac{e^{2}}{\hbar}$ at the TQPT point.}
\end{figure}

More importantly, when considering fluctuation effect in next section, we will find that the mentioned topological quantum phase transition is in fact described by the three dimensional quantum electrodynamics, namely the so-called QED-3, where Dirac fermions are coupled to the dynamic U(1) gauge-field. It is noted that this interesting feature is in contrast to our previous study reported in Ref.\onlinecite{Zhong2013}. The main findings in this subsection are systematically summarized in Fig.~\ref{fig:2}.

\section{Fluctuation effect beyond the mean-field treatment for antiferromagnetic spin-density wave states}\label{sec4}

\subsection{fluctuation effect in antiferromagnetic spin-density wave states}
Here, we discuss the spin fluctuation effect in the antiferromagnetic SDW states, which is neglected in the previous mean-field treatment in Sec. \ref{sec2}.

Firstly, let us recall the effective Dirac action [Eq. (\ref{eq10})]
\begin{eqnarray}
S=\int d^{2}xd\tau\sum_{a\sigma}\left[\bar{\psi}_{a\sigma}(\gamma_{\mu}(\partial_{\mu}-iA_{\mu}^{c}-i\frac{\sigma}{2}A_{\mu}^{s})+m_{a\sigma})\psi_{a\sigma}\right],\nonumber
\end{eqnarray}
where we have defined the effective mass as $m_{1\uparrow}=m-J_{\parallel}/4$, $m_{2\downarrow}=-m-J_{\parallel}/4$, $m_{1\downarrow}=-m+J_{\parallel}/4$ and $m_{2\uparrow}=m+J_{\parallel}/4$ with $m=-3\sqrt{3}t'$.

Following Ref. \onlinecite{He2011}, when the antiferromagnetic order is well established, the effect of spin fluctuations is equivalent to introduce an effective U(1) dynamic gauge-field $b_{\mu}$ into the above action.
In fact, the dynamic U(1) gauge-field $b_{\mu}$ represents the usual spin-wave modes (the Goldstone mode of the antiferromagnetic SDW background) since it has a linear spectrum if we integrate out the Dirac fermions and set
external fields $A_{\mu}^{c}=A_{\mu}^{s}=0$. This point will be clearly seen in discussion below.
\begin{equation}
S'=\int d^{2}xd\tau\sum_{a\sigma}\bar{\psi}_{a\sigma}[\gamma_{\mu}(\partial_{\mu}-iA_{\mu}^{c}-i\frac{\sigma}{2}A_{\mu}^{s}-i\frac{\sigma}{2} b_{\mu})
+m_{a\sigma}]\psi_{a\sigma}.\nonumber
\end{equation}
It is noted that different spin flavor of electrons have the opposite gauge charge of $b_{\mu}$, which just indicates that the gauge-field $b_{\mu}$ describes the spin fluctuations of ordered magnetic background.

After integrating out Dirac fermions, one can obtain
\begin{eqnarray}
&& S'_{\text{CS}}=\sum_{a\sigma}\int d^{2}xd\tau\left[\frac{-ie^{2}m_{a\sigma}}{8\pi|m_{a\sigma}|}\epsilon^{\mu\nu\lambda}(A_{\mu}^{c}+\frac{\sigma}{2}A_{\mu}^{s}+\frac{\sigma}{2} b_{\mu})\right.\nonumber\\
&&\hspace{1cm} \left.\times\partial_{\nu}(A_{\lambda}^{c}+\frac{\sigma}{2}A_{\lambda}^{s}+\frac{\sigma}{2} b_{\lambda})\right].\nonumber
\end{eqnarray}

For N-SDW state, we have $m_{1\uparrow},m_{2\downarrow}<0$ and $m_{1\downarrow},m_{2\uparrow}>0$, thus the resulting action vanishes. Oppositely, if we consider the T-SDW states, ($m_{1\uparrow},m_{2\uparrow}<0$ and $m_{1\downarrow},m_{2\downarrow}>0$) the effective action reads
\begin{equation}
S'_{\text{CS}}=\int d^{2}xd\tau\left[\frac{2i}{4\pi}\epsilon^{\mu\nu\lambda}A_{\mu}^{c}\partial_{\nu}(A_{\lambda}^{s}+b_{\lambda})+\frac{1}{2g}(\epsilon^{\mu\nu\lambda}\partial_{\nu}b_{\lambda})^{2}\right].\nonumber
\end{equation}

Here the Maxwell term for $b_{\mu}$ is reintroduced. Now, if we just turn off the external field $A_{\mu}^{c}$, we get only the Maxwell term for $b_{\mu}$, which clearly denotes a linear spectrum. Since $b_{\mu}$ is the dynamical field, one must integrate it out to obtain the physical classical action. Integrating $b_{\mu}$ out, one has the following action
\begin{equation}
S'_{\text{CS}}=\int d^{2}xd\tau\left[\frac{2i}{4\pi}\epsilon^{\mu\nu\lambda}A_{\mu}^{c}\partial_{\nu}A_{\lambda}^{s}+m_{c}^{2}(A_{\mu}^{c})^{2}\right],\nonumber
\end{equation}
where the external electromagnetic field $A_{\mu}^{c}$ acquires a mass term $m_{c}^{2}\propto g$, which represents the quantum fluctuation (spin-wave) correction in the T-SDW state.
We should emphasize that although there exists an extra mass term for the external electromagnetic field, the spin Hall conductance is still unchanged since it depends the response to the spin gauge-field $A_{\mu}^{s}$.
($J_{x}^{s}=\sigma_{SH}E_{y}=\frac{\partial S_{CS}}{\partial A_{x}^{s}}|_{\vec{A}^{s/c}\rightarrow0}=\frac{2i}{4\pi}(\partial_{y}A_{0}-\partial_{0}A_{y})=\frac{2}{4\pi}E_{y}$.)
Therefore, we may conclude that the spin fluctuation in the antiferromagnetic SDW states does not lead to any qualitative changes on the mean-field results on the previous section.

Some careful readers may wonder that since the external electromagnetic field $A_{\mu}^{c}$ acquires a mass term, does it indicate certain condensation of bosonic objects?
We think that the since the antiferromagnetic SDW states are just the result of the condensation of electrons in the particle-hole channel and the spin-wave or the gauge-field $b_{\mu}$ is the Goldstone mode of such condensation, the mentioned mass term can be considered as a result of the Anderson-Higgs mechanism, which means the electromagnetic field $A_{\mu}^{c}$ kills the gapless Goldstone mode and acquires a mass term.

\subsection{QED-3 emerges as a description of the fluctuation effect at the topological quantum phase transition point}
In this subsection, we discuss the spin fluctuation effect at the topological quantum phase transition point between the N- and the T-SDW states ($m=-3\sqrt{3}t'=-J_{\parallel}/4$) in Sec. \ref{sec2}. Here, the effective mass of Dirac action is $m_{1\uparrow}=-m_{1\downarrow}=-J_{\parallel}/2<0$ and $m_{2\uparrow}=m_{2\downarrow}=0$. Thus, $\psi_{1\uparrow},\psi_{2\downarrow}$ are massive and can be safely integrated out while one should not integrate out $\psi_{2\uparrow},\psi_{2\downarrow}$ due to gaplessness of them. Then, following the treatment of last subsection, one obtains
\begin{eqnarray}
&&S''=S_{\text{CS}}+S_{2\uparrow}+S_{2\downarrow},\nonumber\\
&&S_{\text{CS}}=\int d^{2}xd\tau\left[\frac{i}{4\pi}\epsilon^{\mu\nu\lambda}A_{\mu}^{c}\partial_{\nu}(A_{\lambda}^{s}+b_{\lambda})+\frac{1}{2g}(\epsilon^{\mu\nu\lambda}\partial_{\nu}b_{\lambda})^{2}\right],\nonumber\\
&&S_{2\uparrow}=\int d^{2}xd\tau\left[\bar{\psi}_{2\uparrow}\gamma_{\mu}(\partial_{\mu}-iA_{\mu}^{c}-\frac{i}{2}(A_{\mu}^{s}+b_{\mu}))\psi_{2\uparrow}\right],\nonumber\\
&&S_{2\downarrow}=\int d^{2}xd\tau\left[\bar{\psi}_{2\downarrow}\gamma_{\mu}(\partial_{\mu}-iA_{\mu}^{c}+\frac{i}{2}(A_{\mu}^{s}+b_{\mu}))\psi_{2\downarrow}\right],\nonumber
\end{eqnarray}
where we have integrated out two massive modes, which leads to a mutual Chern-Simons term for $A_{\mu}^{c},A_{\mu}^{s}$ and $b_{\mu}$.
If one is not interested in the external field response of $A_{\mu}^{c}$ and $A_{\mu}^{s}$, we may neglect them and get the effective action as
\begin{eqnarray}
&&S''=\int d^{2}xd\tau\left[\frac{1}{2g}(\epsilon^{\mu\nu\lambda}\partial_{\nu}b_{\lambda})^{2}+\sum_{\sigma}\bar{\psi}_{2\sigma}\gamma_{\mu}(\partial_{\mu}-\frac{i}{2}\sigma b_{\mu})\psi_{2\sigma}\right]\nonumber
\end{eqnarray}

Interestingly, this action is just the well-known three dimensional quantum electrodynamics (QED-3) but with opposite coupling for different flavor Dirac fermions. The QED-3 action appears in the gauge-field description of
high temperature superconductivity and some quantum magnetism problems.\cite{Wen} But, in those gauge-field description, the Dirac fermions are not the physical electrons so the Green's function of the Dirac fermions are not physical observable while in our case, the Dirac fermions are real electrons, thus it indeed describes true physics. In other words, if we obtain the Green's function of the Dirac fermions in our model, one can get its corresponding local density of state, which could be measured by local differential conductance in scanning tunneling microscopy experiments.

Unfortunately, for the QED-3, we do not know the exact results for the
gauge-field and fermion Green's functions. And here, we only present the one-loop/large-N result.\cite{Franz2002}
Integrating out Dirac fermions at the one-loop level, we can get the action for the gauge-field $b_{\mu}$,
\begin{eqnarray}
S_{b}=\int\frac{d^{3}q}{(2\pi)^{3}}b_{\mu}\frac{\sqrt{q^{2}}}{32}\left(\delta_{\mu\nu}-\frac{q_{\mu}q_{\nu}}{q^{2}}\right)b_{\nu}.\nonumber
\end{eqnarray}
So the Green's function for the gauge-field is $D_{\mu\nu}(q)=\langle b_{\mu}(q)b_{\nu}(-q)\rangle=\frac{16}{\sqrt{q^{2}}}\left(\delta_{\mu\nu}-\frac{q_{\mu}q_{\nu}}{q^{2}}\right)$.
Then, using this Green's function, one can compute the self-energy correction for the Dirac fermions,

\begin{eqnarray}
\Sigma(k)=\int\frac{d^{3}q}{(2\pi)^{3}}D_{\mu\nu}(q)\gamma_{\mu}G_{0}(k+q)\gamma_{\nu},\nonumber
\end{eqnarray}
where $G_{0}(k)=\frac{ik_{\alpha}\gamma_{\alpha}}{k^{2}}$ is the Green's function for free Dirac fermions. Calculating this integral (for details, one can refer to Appendix \ref{C}), we find
$\Sigma(k)\simeq \frac{16i}{3\pi^{2}}\gamma_{\mu}k_{\mu}\ln(\frac{\Lambda}{|k|})$ and the Dirac fermion Green's function reads
$G(k)=(G_{0}(k)^{-1}-\Sigma(k))^{-1}=(-ik_{\alpha}\gamma_{\alpha}(1+\frac{16}{3\pi^{2}}\ln(\frac{\Lambda}{|k|}))^{-1}\simeq(-ik_{\alpha}\gamma_{\alpha}e^{\frac{16}{3\pi^{2}}\ln(\frac{\Lambda}{|k|})})^{-1}=\frac{ik_{\alpha}\gamma_{\alpha}}{k^{2-\eta}\Lambda^{\eta}}$
with $\eta=\frac{16}{3\pi^{2}}$. Superficially, this indicates a rather large anomalous dimension $\eta\simeq 0.54$ for Dirac fermions, but the one-loop result is not reliable and the realistic situation
is not clear at present stage. Since we do not have a good understanding of the QED-3 theory, the present topological quantum phase transition point is obviously poorly understood and this interesting issuer is left for future study.

One may wonder that since the gauge-field $b_{\mu}$ does not have its own Chern-Simons term, the action of the QED-3 may be meaningless due to the well-known confinement.\cite{Polyakov1975,Polyakov1977}
However, we should note that the gauge-field $b_{\mu}$ is in fact coupled to the physical external electromagnetic field $A_{\mu}^{c}$ and since the $A_{\mu}^{c}$ is obviously deconfined in the three-dimensional time-space, the
gauge-field $b_{\mu}$ has to be deconfined in the present case. In other words, deconfinement of $b_{\mu}$ is protected by its coupling to the physical external electromagnetic field.
Similar case also appears in the study of the quantum phase transition of bosonic integer quantum Hall phases.\cite{Grover2012,Lu2012b}

After all, the topological quantum phase transition between the T and N-SDW states can be described by an effective QED-3 theory though understanding exactly physics of this theory is beyond our ability.

\section{The Kondo insulating state, quantum spin Hall insulator and Kondo breakdown }\label{sec5}
Having discussed the properties of magnetic ordered states in the previous sections, it is time to study paramagnetic states of the Kane-Mele-Kondo lattice model in order to compare with the previous results.
The paramagnetic states include the Kondo insulator and the quantum spin Hall insulator, which are also discussed in Ref. \onlinecite{Feng}. So, we will give a brief discussion on these two states
based on our mean-field Hamiltonian Eq. (\ref{eq2}). The basic physical features are consistent with the previous work.\cite{Feng}

\subsection{The Kondo insulating state}
Another interesting case appears when $J_{\parallel}\ll J_{\perp}$. It is natural to expect that a Kondo insulating state arises in this situation for half-filling.\cite{Tsunetsugu,Zhang2000} Following the same methology of treating antiferromagnetic SDW state, we can get the mean-field Hamintonian and its ground-state energy per site for the expected Kondo insulating state with $V\neq0$ but no magnetic orders $m_{d}=m_{c}=0$
\begin{eqnarray}
&&H_{\text{MF}}=H_{\text{K}}+H_{\perp}^{\text{MF}}+E_{0},\nonumber\\
&&H_{\perp}^{\text{MF}}=\frac{J_{\perp}V}{2}\sum_{k\sigma}(c_{kA\sigma}^{\dag}d_{kA\sigma}+c_{kB\sigma}^{\dag}d_{kB\sigma}+h.c.),\nonumber\\
&&E_{0}=N_{s}J_{\perp}V^{2}.\nonumber
\end{eqnarray}

\begin{eqnarray}
E_{g}^{Kondo}&&=J_{\perp}V^{2}-\frac{4}{3\Lambda^{2}}[(\Lambda^{2}+(3\sqrt{3}t')^{2}+J_{\perp}^{2}V^{2})^{3/2}\nonumber\\
&&-((3\sqrt{3}t')^{2}+J_{\perp}^{2}V^{2})^{3/2}].\label{eq14}
\end{eqnarray}

And the four quasi-particle energy bands are
\begin{eqnarray}
E_{\pm\pm\sigma}(k)&&=\pm\frac{1}{2}[\sqrt{t^{2}|f(k)|^{2}+\gamma(k)^{2}+(J_{\bot}V)^{2}}\nonumber\\
&&\pm\sqrt{t^{2}|f(k)|^{2}+\gamma(k)^{2}}].\label{eq15}
\end{eqnarray}
Minimizing $E_{g}^{Kondo}$ with respect to the Kondo hybridization parameter $V$, we can obtain $V^{2}=\frac{1}{16J_{\perp}^{4}}[\Lambda^{4}-8\Lambda^{2}J_{\perp}^{2}-16J_{\perp}^{2}(3\sqrt{3}t')^{2}+16J_{\perp}^{4}]$, which implies a critical coupling $J_{\perp}^{c}=\frac{2}{\Lambda^{2}}[\sqrt{\Lambda^{2}+(3\sqrt{3}t')^{2}}-3\sqrt{3}t']$ corresponding to the vanishing $V$. The behavior $V\propto(J_{\perp}-J_{\perp}^{c})$ is in contrast to usual mean-field result $\beta=1/2$, and this is due to the low-energy linear DOS of conduction electrons near Dirac points of the honeycomb lattice.

\subsection{The quantum spin Hall insulator and Kondo breakdown}
In fact, the existence of the critical coupling $J_{\perp}^{c}$ results from the competition between the Kondo insulating state and the quantum spin Hall insulator.
(This issue is studied in details in Ref.\onlinecite{Feng}.) The mentioned quantum spin Hall state is defined by $V=m_{d}=m_{c}=0$ at the mean-field level, where the local spins are fully decoupled from the conduction electrons
and the conduction electrons are described by the pure Kane-Mele model Eq. (\ref{eq3}).
\begin{eqnarray}
H_{\text{K}}=-t\sum_{\langle ij\rangle \sigma}c_{i\sigma}^{\dag}c_{j\sigma}-t'\sum_{\langle\langle ij\rangle\rangle \sigma}\sigma e^{i\varphi_{ij}}c_{i\sigma}^{\dag}c_{j\sigma}.\nonumber
\end{eqnarray}
For this single-particle Hamiltonian, its ground-state energy is readily to calculate as $E_{g}^{0}=-\frac{4}{3\Lambda^{2}}[(\Lambda^{2}+(3\sqrt{3}t')^{2})^{3/2}-(3\sqrt{3}t')^{3}]$ which comes solely from free conduction electrons. Comparing $E_{g}^{0}$ and $E_{g}^{Kondo}$, one clearly recovers the critical coupling $J_{\perp}^{c}$, which justifies the above simple picture.

Since the Kondo insulating state is unstable to the QSH state when $J_{\perp}<J_{\perp}^{c}$, one may wonder whether the QSH state appears between the Kondo insulating state and the antiferromagnetic SDW states.
Particularly, if the QSH state is stable, there will exist a critical point corresponds to the Kondo breakdown\cite{Senthil2003,Senthil2004,Pepin2005,Vojta,Rosch} on the honeycomb lattice when $J_{\perp}=J_{\perp}^{c}$, which has been studied in Refs.\onlinecite{Feng} and \onlinecite{Saremi}. Actually, the authors in Ref.\onlinecite{Saremi} also discusses the U(1) gauge fluctuation around the above mean-field ground-state and found that the critical theory for this putative Kondo breakdown point is described by
multi-flavor Dirac fermions and an O(2) boson field coupled to a dynamic U(1) gauge field. The details can be consulted that reference.
\begin{figure}
\includegraphics[width=0.8\columnwidth]{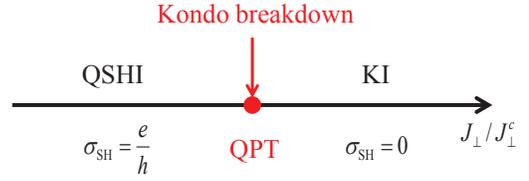}
\caption{\label{fig:3} The ground-state phase diagram of the Kane-Mele Kondo lattice model with only paramagnetic phases considering. KI is the Kondo insulator and QSHI denotes the quantum spin Hall insulator. The quantum phase transition between the KI and QSHI is second order and is considered as a Kondo breakdown transition. However, when magnetic orders is introduced, QSHI is unstable and this QPT is meaningless.}
\end{figure}

However, recalling of the ground-state energy of the antiferromagnetic SDW state $E_{g}^{AFM}$, we find that it is always lower than $E_{g}^{0}$ for any positive Kondo coupling $J_{\parallel}$ and the next-nearest-neighbor hopping $t'$. Therefore, we conclude that the nontrivial quantum spin Hall state in intermediate coupling seems unfavorable based on our present mean-field treatment and the stability of QSH state in Ref.\onlinecite{Feng} may require
extra frustration interactions or certain geometrical frustrations.\cite{Coleman2010,Custers2010} (For example, a $J_{1}-J_{2}$ Heisenberg exchange term for local spins may be added into the Kondo lattice model and if
appropriately tuning the ratio of $J_{1}/J_{2}$ may frustrate the magnetic ordered states and lead to the wanted paramagnetic QSH phase.\cite{Sachdev2011}) In other words, we only expect the system supports the phases of SDW states, the paramagnetic Kondo insulator state and the phase transitions between them.

\subsection{Fluctuation in Kondo insulator and QSH insulator}
Since both the Kondo insulator and QSH state are gapped, the small fluctuation cannot change the basic physics of those states. To be specific, for the Kondo insulator, the fluctuation will mainly come from the fluctuation of Kondo hybridizing part $-V=\langle c_{i\uparrow}^{\dag}d_{i\uparrow}+d_{i\downarrow}^{\dag}c_{i\downarrow}\rangle =
\langle c_{i\downarrow}^{\dag}d_{i\downarrow}+d_{i\uparrow}^{\dag}c_{i\uparrow}\rangle$ if we treat $V$ as a dynamic complex bosonic field. However, since $V$ acquires nonzero expectation in the
Kondo insulator phase, the possible gapless fluctuation of it is its phase degree of freedom. However, such phase degree of freedom can be absorbed into the gapped Lagrangian multiplier field.\cite{Read} So,
$V$ cannot give rise to any dangerous gapless fluctuation and thus the Kondo insulator is stable.

As for the QSH insulator, unless the fluctuation due to the Kondo hybridizing and magnetic order terms is comparable to the single-particle gap of the QSH state, we still expect the QSH state to be stable to fluctuations.

\section{Constructing the ground-state phase diagram of Kane-Mele-Kondo lattice} \label{sec6}
\begin{figure}
\includegraphics[width=0.8\columnwidth]{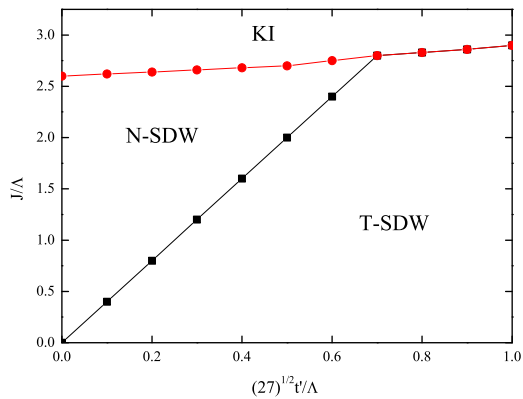}
\caption{\label{fig:4} The ground-state phase diagram of the Kane-Mele-Kondo lattice model. The KI is the Kondo insulator and the quantum phase transition between the KI and the antiferromagnetic SDW states is first order. In the antiferromagnetic SDW regime, N-SDW denotes the normal antiferromagnetic SDW state while the T-SDW represents the topological antiferromagnetic SDW state with a quantum spin Hall conductance and helical edge-states. The boundary of these two kinds of SDW states is the topological quantum phase transition, which is determined by condition of $3\sqrt{3}t'=J/4$ with $\Lambda\simeq2.33t$ being high-energy cutoff.}
\end{figure}

Based on the previous sections, in which the T-SDW, N-SDW, QSH insulator and Kondo insulator states are individually discussed, in Fig. \ref{fig:4} we show the global ground-state phase diagram. Here we do not consider possible coexistence regime for simplicity. In this phase diagram, the QSH insulator is absent due to its energy is always higher than that of antiferromagnetic SDW states, as discussed in last section. The Kondo insulator is the same as that in Ref. \onlinecite{Feng}. In addition, the two kinds of antiferromagnetic SDW states are central results of the present work and the transition between them is topological quantum phase transition determined by the condition of close of gap $3\sqrt{3}t'=J/4$. The phase boundary between two antiferromagnetic SDW states and the Kondo insulator is found to be first-order. \cite{Continentino}

\section{Further discussions and possible extensions} \label{sec7}

\subsection{Doping case: Possibility of a nontrivial topological phase}
The above Sections have focused on the half-filled Kane-Mele-Kondo lattice model since in this case, one can expect nontrivial topological properties when the spin Chern number is nonzero due to full occupation of certain single-particle energy bands. However, if the system is doped, it is difficult to expect nontrivial topological state like QSH insulator or the T-SDW state studied in the present paper. However, one notes that the authors in Ref.\onlinecite{Feng} found that the 1/4 and 3/4 filling cases also have nontrivial $Z_{2}$ topological bulk invariant and their surface-state is a new helical edge-state, which contains the contribution from both conduction and local electrons (spins). (Recalling that the usual helical edge-state only includes the contribution of conduction electrons and this appears in the QSH and T-SDW states, which we have studied in the main text.)

\begin{figure}
\includegraphics[width=0.6\columnwidth]{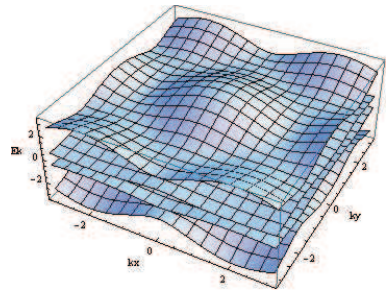}
\caption{\label{fig:5} The four quasiparticle energy bands in the Kondo insulator phase with $t=1,t'=0.2,J_{\bot}=3,V=0.77$ [See Eq. (\ref{eq15}) for detailed formalism.]}
\end{figure}

To get more insight into the mentioned situation of the 1/4 and 3/4 filling, we plot the quasi-particle energy bands in Fig.\ref{fig:5}.
We can see that when the 1/4 filling is considered, the lowest band is filled and any single-particle excitation to the other bands has a gap. The 3/4 filling case is the same to 1/4 filling. However,
we cannot use the formula $C_{1}^{\sigma}=\frac{1}{4\pi}\int dk_{x}dk_{y}\hat{\textbf{d}_{\sigma}}\cdot(\frac{\partial\hat{\textbf{d}_{\sigma}}}{\partial k_{x}}\times \frac{\partial\hat{\textbf{d}_{\sigma}}}{\partial k_{y}})$\cite{DSheng2006,Prodan2009,Qi2011} to calculate the spin Chern number, thus it is not straightforward to construct a corresponding Chern-Simons-like effective theory. It is interesting to study this point in the future
by developing other effective theories to capture the nontrivial physics at 1/4 and 3/4 filling.

\subsection{Doping the Kane-Mele-Kondo lattice: Relevance of Ferromagnetism}
In usual Kondo lattice model, when the conduction electron has small filling fraction, the ferromagnetism may be relevant and could occupy some regime of the ground-state phase diagram.\cite{Tsunetsugu,Zhang2010}
This means that at small filling, the antiferromagnetic states are unstable to ferromagnetic states. This may indicate that when one dopes the Kane-Mele-Kondo lattice strongly deviating the half-filling case,
one could find a ferromagnetic phase whose mean-field Hamiltonian reads
\begin{eqnarray}
&&H_{\text{MF}}=H_{\text{c}}+H_{\text{d}}+E_{0},\nonumber\\
&&H_{\text{c}}=\sum_{k\sigma}-t[f(k)c_{kA\sigma}^{\dag}c_{kB\sigma}+f^{\star}(k)c_{kB\sigma}^{\dag}c_{kA\sigma}] \nonumber\\
&& + (2t'\sigma\gamma(k)+\frac{J_{\parallel}}{2}\sigma m_{d}-\mu)(c_{kA\sigma}^{\dag}c_{kA\sigma}+c_{kB\sigma}^{\dag}c_{kB\sigma}),\nonumber\\
&&H_{\text{d}}=\sum_{k\sigma}\left[(-m_{c}\frac{J_{\parallel}}{2})\sigma+\lambda\right](d_{kA\sigma}^{\dag}d_{kA\sigma}+d_{kB\sigma}^{\dag}d_{kB\sigma}),\nonumber\\
&&E_{0}=N_{s}(2J_{\parallel}m_{d}m_{c}-\lambda), \nonumber
\end{eqnarray}
where we have defined the mean-field parameters $\langle d_{i\uparrow}^{\dag}d_{i\uparrow}-d_{i\downarrow}^{\dag}d_{i\downarrow}\rangle = 2m_{d}$, $\langle c_{i\uparrow}^{\dag}c_{i\uparrow}-c_{i\downarrow}^{\dag}c_{i\downarrow}\rangle = -2m_{c}$. (Those parameters are the same for A and B sublattice.) The Lagrangian multiplier $\lambda$ is introduced to enforce the single-occupation of local fermions and the chemical potential $\mu$ is reintroduced.

Computing the ground-state energy of ferromagnetic state, one may compare it with the energy of antiferromagnetic states and Kondo states (the heavy fermion liquid) to establish the ground-state phase diagram.
However, we do not expect nontrivial ferromagnetic state with quantized spin Hall conductance since the system deviates from the half-filling situation.

\subsection{Remarks on the topological Kondo insulator and its Anderson lattice model}
In addition, we note that in literature, the so-called topological
Kondo insulator has been proposed as an extension of the original
topological (band) insulators.\cite{Dzero} In contrast to usual topological
insulators where the spin-orbit coupling is encoded in a
spin-dependent hopping amplitudes between different unit cells, the
topological Kondo insulator is produced by the spin-orbit coupling
associated with the hybridization between conduction and local
electrons.\cite{Dzero,Dzero2012,Kim2012} In their papers, the Anderson lattice model is used with
nontrivial hybridization between conduction and local electrons. However, due to the nontrivial hybridization, it is not clear whether one could derive
a corresponding low-energy Kondo lattice-like model in the Kondo limit. Therefore, it is not easy to compare the results of the present Kane-Mele-Kondo lattice model to their Anderson lattice model.

On the other hand, there exists numerical studies on the proposed Anderson lattice model.\cite{Yoshida,Assaad2013} The topological Kondo insulator
is confirmed by those dynamical mean-field theory calculations. Interestingly, the authors in Ref.\onlinecite{Yoshida} report a ferromagnetic state with
a nonzero spin Chern number, which seems a new topological state like the T-SDW state found in the present paper, namely a state with both the magnetic order and a quantized charge or spin Chern number/Hall conductance.

\subsection{Possible directions for numerical study}
To our knowledge, there does not exist any numerical studies on the Kane-Mele-Kondo lattice or the Kane-Mele-Anderson lattice model up to now. The Kane-Mele-Kondo lattice model is defined by
Eq. (\ref{eq1}) and the Kane-Mele-Anderson lattice model is defined as\cite{Feng}
\begin{eqnarray}
&&H=H_{\text{K}}+H_{\text{V}}+H_{\text{d}},\nonumber\\
&&H_{\text{K}}=-t\sum_{\langle ij\rangle \sigma}c_{i\sigma}^{\dag}c_{j\sigma}-t'\sum_{\langle\langle ij\rangle\rangle \sigma}\sigma e^{i\varphi_{ij}}c_{i\sigma}^{\dag}c_{j\sigma},\nonumber\\
&&H_{\text{V}}=V\sum_{i\sigma}(c_{i\sigma}^{\dag}d_{i\sigma}+d_{i\sigma}^{\dag}c_{i\sigma}),\nonumber\\
&&H_{\text{d}}=\sum_{i\sigma}\varepsilon_{d}d_{i\sigma}^{\dag}d_{i\sigma}+U\sum_{i}d_{i\uparrow}^{\dag}d_{i\uparrow}d_{i\downarrow}^{\dag}d_{i\downarrow},\nonumber
\end{eqnarray}
where $V$ is the hybridization between conduction electrons $c_{\sigma}$ and local electrons $d_{\sigma}$, which is a free parameter in the Anderson lattice model.
$\varepsilon_{d}$ denotes the local energy level of local electrons and $U$ is the usual local Hubbard repulsive energy, which represents the local strong correlation feature of local electrons.

In literature, the Kane-Mele model with Hubbard-$U$ interaction, i.e., the Kane-Mele-Hubbard model, is widely studed by various numerical techniques such as auxiliary field quantum Monte Carlo,\cite{Hohenadler2011,Zheng2011} variational cluster,\cite{Li2011} variational Monte Carlo\cite{Imada2011} and cellular dynamical mean-field theory. \cite{Wu2012}
We expect that since the Kane-Mele-Anderson lattice is not very different from the Kane-Mele-Hubbard model, those numerical techniques may be utilized to capture the ground-state phase diagram and many correlation functions
though the Kane-Mele-Anderson lattice has larger Hilbert space than the Kane-Mele-Hubbard model (Note that there exist two kinds of electrons, the conduction and local electrons in Kane-Mele-Anderson lattice while only conduction
electrons appear in the Kane-Mele-Hubbard model).

\subsection{Implication for experiments and realization of the Kane-Mele-Kondo model}
Recently, the Kondo insulator samarium hexaboride SmB$_{6}$, which is proposed as a promising candidate for the topological Kondo insulator, is hotly
studied by many experimental groups.\cite{NXu2013,Neupane2013,Li2013} We note that SmB$_{6}$ is basically a three-dimensional material, thus our theoretical results cannot be directly applied to this interesting example. Materials relevant to our theoretical results should be certain quasi-two-dimensional compounds with strong spin-orbit coupling and strong local interactions. Particularly, if their lattice is the honeycomb lattice, it will be a good chance to compare the experimental results to the theoretical predictions, as shown in Fig. \ref{fig:4}.

Since there have no realistic materials, which could be well modeled by the Kane-Mele-Kondo lattice model, we have to expect the versatile ultra-cold atoms systems may realize our proposed model on the honeycomb lattice.\cite{Bloch,Goldman,Shao}

\subsection{Three-dimensional system with interplay of magnetism and topology}
Till now, the theoretical study of the topological Kondo states is only focused on the topological Kondo insulator though
we know that even the most promising topological Kondo insulator material SmB$_{6}$ can also show an antiferromagnetic state under external pressure.
Thus, it is interesting to theoretically discuss the transition from the putative topological Kondo insulator to the antiferromagnetic ordered state
in certain three-dimensional lattice models. Recalling that the topological Kondo insulator is described by the Anderson lattice models in Refs.\onlinecite{Dzero} and \onlinecite{Kim2012}, introducing a Heisenberg exchange interaction between the spin degree of freedom of local spins may include the effect of magnetism into the original models.

\section{conclusion} \label{sec8}

In summary, we have obtained the ground-state phase diagram of the Kane-Mele-Kondo lattice model on the honeycomb lattice at half-filling. It is found that the paramagnetic Kondo insulator and the magnetic normal and topological antiferromagnetic spin-density-wave states occupy the regimes of the phase diagram while the quantum spin Hall insulator is excluded. The topological magnetic state has the same symmetry as the normal one but it has an extra quantized spin Hall conductance and a nontrivial helical edge-state. Interestingly, those topological properties can be captured in terms of a mutual Chern-Simons theory and the stability of the helical edge-state is protected by recovered time-reversal symmetry. Furthermore, there exists a novel topological quantum phase transition between the two kinds of magnetic phases, whose critical behaviors may be described by a three-dimensional quantum electrodynamics.
In addition, a first-order quantum phase transition is found between the Kondo insulating state and the antiferromagnetic states. We hope the present work may be helpful for further studies on the interplay between conduction electrons and the densely localized spins on the honeycomb lattice.

\begin{acknowledgments}
The work was supported partly by NSFC, PCSIRT (Grant No. IRT1251), the Program for NCET, the Fundamental Research Funds for the Central Universities and the national program for basic research of China.
\end{acknowledgments}

\appendix
\section{Derivation of Chern-Simons action}\label{A}
Here, we would like to give a brief derivation of the effective Chern-Simons action Eq. (\ref{eq6}) from the Dirac action Eq. (\ref{eq5}) but without the spin gauge-field $A_{\mu}^{s}$. Introducing the spin gauge-field
is easy to implement as just a replacement ($A_{\mu}^{c}\rightarrow A_{\mu}^{c}+\frac{\sigma}{2e}A_{\mu}^{s}$).

Firstly, we consider the following Dirac action
\begin{equation}
S=\int d^{2}xd\tau[\bar{\psi}(\gamma_{\mu}(\partial_{\mu}-ieA_{\mu}^{c})+m)\psi].
\end{equation}

Then, integrating out Dirac fermions one obtains
\begin{eqnarray}
&& S_{eff}=\ln \text{Det}[\gamma_{\mu}(\partial_{\mu}-ieA_{\mu}^{c})+m]\nonumber\\
&& =\text{Tr}\ln[\gamma_{\mu}(\partial_{\mu}-ieA_{\mu}^{c})+m]\nonumber\\
&& =\text{Tr}[\ln[\gamma_{\mu}\partial_{\mu}+m]+\ln[1-ie(\gamma_{\mu}\partial_{\mu}+m)^{-1}\gamma_{\mu}A_{\mu}^{c}]]\nonumber\\
&& \simeq \int\frac{d^{3}q}{(2\pi)^{3}}A_{\mu}^{c}\Pi_{\mu\nu}(q)A_{\nu}^{c}
\end{eqnarray}
where $\Pi_{\mu\nu}(q)=\frac{-e^{2}}{2}\int\frac{d^{3}k}{(2\pi)^{3}}\text{Tr}[\frac{ik_{\alpha}\gamma_{\alpha}-m}{m^{2}+k^{2}}\gamma_{\nu}\frac{i(k_{\beta}+q_{\beta})\gamma_{\beta}-m}{m^{2}+(k+q)^{2}}\gamma_{\mu}]
=\frac{-e^{2}}{2}\int\frac{d^{3}k}{(2\pi)^{3}}[\frac{1}{m^{2}+k^{2}}\frac{1}{m^{2}+(k+q)^{2}}][-imq_{\lambda}Tr(\gamma_{\mu}\gamma_{\nu}\gamma_{\lambda})]+...
\simeq\frac{-e^{2}m}{8\pi|m|}\epsilon^{\mu\nu\lambda}q_{\lambda}$ is calculated at one-loop level. We have also used the identity $\text{Tr}(\gamma_{\mu}\gamma_{\nu}\gamma_{\lambda})=2i\epsilon^{\mu\nu\lambda}$ with $\gamma_{0}=\tau_{z}$, $\gamma_{1}=\tau_{x}$ and $\gamma_{2}=\tau_{y}$ while $\int\frac{d^{3}k}{(2\pi)^{3}}[\frac{1}{m^{2}+k^{2}}\frac{1}{m^{2}+(k+q)^{2}}]=\frac{\arcsin\left(\frac{|q|}{\sqrt{q^{2}+4m^{2}}}\right)}{4\pi|q|}\simeq\frac{1}{8\pi|m|}$ for $|q|\ll|m|$. Therefore, the effective Chern-Simon action is obtained as
\begin{eqnarray}
S_{eff}&&=\int\frac{d^{3}q}{(2\pi)^{3}} A_{\mu}^{c}\frac{-e^{2}m}{8\pi|m|}\epsilon^{\mu\nu\lambda}q_{\lambda}A_{\nu}^{c}\nonumber\\
&&=\int d^{2}xd\tau[e^{2}\frac{-i m}{8\pi|m|}\epsilon^{\mu\nu\lambda}A_{\mu}^{c}\partial_{\nu}A_{\lambda}^{c}].
\end{eqnarray}

Consider Eq. (\ref{eq5}), we have four-flavor Dirac fields and the effective action reads
\begin{eqnarray}
S_{eff}&&=\int d^{2}xd\tau\sum_{a\sigma}e^{2}\frac{-i m_{a\sigma}}{8\pi|m_{a\sigma}|}\epsilon^{\mu\nu\lambda}(A_{\mu}^{c}+\frac{\sigma}{2e}A_{\mu}^{s})\nonumber\\
&&\times\partial_{\nu}(A_{\lambda}^{c}+\frac{\sigma}{2e}A_{\lambda}^{s})].\nonumber
\end{eqnarray}
Since $m_{1\sigma}=m_{2\sigma}=m\sigma$, one gets
\begin{eqnarray}
S_{eff}&&=\int d^{2}xd\tau\sum_{\sigma}e^{2}\frac{-i m\sigma}{8\pi|m\sigma|}\epsilon^{\mu\nu\lambda}(A_{\mu}^{c}+\frac{\sigma}{2e}A_{\mu}^{s})\nonumber\\
&&\times\partial_{\nu}(A_{\lambda}^{c}+\frac{\sigma}{2e}A_{\lambda}^{s})].\nonumber
\end{eqnarray}
The above action is just Eq. (\ref{eq6}) if $e$ is set to unit.

\section{Derivation of the polarization function}\label{B}
Using the formalism of the polarization function in Appendix. \ref{A} with $m=0$,
\begin{eqnarray}
\Pi_{\mu\nu}(q)&&=\frac{-e^{2}}{2}\int\frac{d^{3}k}{(2\pi)^{3}}\text{Tr}[\frac{ik_{\alpha}\gamma_{\alpha}}{k^{2}}\gamma_{\nu}\frac{i(k_{\beta}+q_{\beta})\gamma_{\beta}}{(k+q)^{2}}\gamma_{\mu}]\nonumber\\
&&=\text{Tr}(\gamma_{\alpha}\gamma_{\nu}\gamma_{\beta}\gamma_{\mu})\frac{e^{2}}{2}\int\frac{d^{3}k}{(2\pi)^{3}}\frac{k_{\alpha}}{k^{2}}\frac{(k_{\beta}+q_{\beta})}{(k+q)^{2}}\nonumber\\
&&=\frac{e^{2}}{16}(\delta_{\mu\nu}-\frac{q_{\mu}q_{\nu}}{q^{2}})\sqrt{q^{2}}.\label{B1}
\end{eqnarray}
where we have used the identity $\text{Tr}(\gamma_{\alpha}\gamma_{\beta})=2\delta_{\alpha\beta}$ and $\text{Tr}(\gamma_{\alpha}\gamma_{\nu}\gamma_{\beta}\gamma_{\mu})=2\delta_{\alpha\nu}\delta_{\beta\mu}-2\delta_{\alpha\beta}\delta_{\mu\nu}+2\delta_{\alpha\mu}\delta_{\beta\nu}$.
The complex integral is calculated as
\begin{eqnarray}
&&\int\frac{d^{3}k}{(2\pi)^{3}}\frac{k_{\alpha}}{k^{2}}\frac{(k_{\beta}+q_{\beta})}{(k+q)^{2}}\nonumber\\
&&=\int_{0}^{1} dx\int\frac{d^{3}k}{(2\pi)^{3}}\frac{k_{\alpha}(k_{\beta}+q_{\beta})}{[k^{2}x+(1-x)(k+q)]^{2}}\nonumber\\
&&=\int_{0}^{1} dx\int\frac{d^{3}k'}{(2\pi)^{3}}\frac{k'_{\alpha}k'_{\beta}+x k'_{\alpha}q_{\beta}+(x-1)k'_{\beta}q_{\alpha}+x(x-1)q_{\alpha}q_{\beta}}{[k'^{2}+x(1-x)q^{2}]^{2}}\nonumber\\
&&=\int_{0}^{1} dx\int\frac{d^{3}k'}{(2\pi)^{3}}\frac{k'_{\alpha}k'_{\beta}+x(x-1)q_{\alpha}q_{\beta}}{[k'^{2}+x(1-x)q^{2}]^{2}}\nonumber\\
&&=\int_{0}^{1} dx\int\frac{d^{3}k'}{(2\pi)^{3}}\frac{\frac{1}{3}\delta_{\alpha\beta}k'^{2}+x(x-1)q_{\alpha}q_{\beta}}{[k'^{2}+x(1-x)q^{2}]^{2}}\nonumber\\
&&=-(\frac{q_{\alpha}q_{\beta}}{q^{2}}+\delta_{\alpha\beta})\sqrt{q^{2}}\Gamma(\frac{1}{2})\frac{1}{4\pi^{3/2}}\int_{0}^{1} dx\sqrt{x(1-x)}\nonumber\\
&&=-\frac{1}{32}(\frac{q_{\alpha}q_{\beta}}{q^{2}}+\delta_{\alpha\beta})\sqrt{q^{2}}\nonumber
\end{eqnarray}
with $\Gamma(\frac{1}{2})=\sqrt{\pi}$ and $\int_{0}^{1} dx\sqrt{x(1-x)}=\pi/8$.

\section{Derivation of the self-energy correction}\label{C}
\begin{eqnarray}
\Sigma(k)&&=\int\frac{d^{3}q}{(2\pi)^{3}}(D_{\mu\nu}(q)\gamma_{\mu}G_{0}(k+q)\gamma_{\nu})\nonumber\\
&&=\int\frac{d^{3}q}{(2\pi)^{3}}\frac{16}{\sqrt{q^{2}}}(\delta_{\mu\nu}-\frac{q_{\mu}q_{\nu}}{q^{2}})\gamma_{\mu}\frac{i(k+q)_{\alpha}\gamma_{\alpha}}{(k+q)^{2}}\gamma_{\nu}.\nonumber
\end{eqnarray}
Then, using the identity $(\delta_{\mu\nu}-\frac{q_{\mu}q_{\nu}}{q^{2}})\gamma_{\mu}\gamma_{\alpha}\gamma_{\nu}=-2q_{\mu}\gamma_{\mu}\frac{q_{\alpha}}{q^{2}}$,
we get
\begin{eqnarray}
\Sigma(k)=\frac{-4i}{\pi^{3}}\gamma_{\mu}I_{\mu}(k)=\frac{-4i}{\pi^{3}}\gamma_{\mu}\int d^{3}q q_{\mu}\frac{q(k+q)}{|q|^{3}(k+q)^{2}}.\nonumber
\end{eqnarray}
with $I_{\mu}(k)\equiv \int d^{3}q q_{\mu}\frac{q(k+q)}{|q|^{3}(k+q)^{2}}=C(k)\frac{k_{\mu}}{k^{2}}$.
For the $C(k)$, one has
\begin{eqnarray}
C(k)=\int d^{3}q \frac{(q\cdot k)(q\cdot k+q^{2})}{|q|^{3}(k+q)^{2}}.\nonumber
\end{eqnarray}
In Ref.\onlinecite{Franz2002}, they calculated $C(k)\equiv k_{\mu}I_{\mu}(k)\simeq-\frac{4\pi}{3}k^{2}\ln(\frac{\Lambda}{|k|})$ with $\Lambda$ being the high-energy cut-off.
Therefore, we obtain $\Sigma(k)\simeq \frac{16i}{3\pi^{2}}\gamma_{\mu}k_{\mu}\ln(\frac{\Lambda}{|k|})$.

\end{document}